\documentclass[12pt]{article}
\usepackage{epsfig}
\usepackage{axodraw}
\usepackage{graphicx}
\usepackage{bm}
\usepackage{setspace}
\usepackage{color}

\newcommand{\la}{{\lambda}}
\newcommand{\ie}{{\it i.e.}}
\newcommand{\eg}{{\it e.g.}}

\newcommand{\be}{\begin{equation}}
\newcommand{\ee}{\end{equation}}
\newcommand{\beq}{\begin{equation}}
\newcommand{\eeq}{\end{equation}}
\newcommand{\bea}{\begin{eqnarray}}
\newcommand{\eea}{\end{eqnarray}}
\newcommand{\br}{\begin{eqnarray}}
\newcommand{\er}{\end{eqnarray}}
\newcommand{\ba}{\begin{array}}
\newcommand{\ea}{\end{array}}
\newcommand{\bi}{\begin{itemize}}
\newcommand{\ei}{\end{itemize}}
\newcommand{\bn}{\begin{enumerate}}
\newcommand{\en}{\end{enumerate}}
\newcommand{\bc}{\begin{center}}
\newcommand{\ec}{\end{center}}

\def\bY{{\bf Y}}
\def\bbe{\bm{\beta}}
\def\bg{\bm{\gamma}}
\def\bkappa{\bm{\kappa}}
\def\bk5{\bm{\kappa}_5}
\def\bD{\bm{\Delta}}
\def\bA{{\bf A}}
\def\bB{{\bf B}}
\def\bC{{\bf C}}
\def\bP{{\bf P}}
\def\bU{{\bf U}}
\def\bV{{\bf V}}
\def\bmm{{\bf m}}

\def\bk{{\cal {\bf K}}}
\def\tl{{\tilde{L}}}

\def\tm{{\tilde{m}}}
\def\te{{\tilde{e^c}}}

\def\td{{\tilde{d^c}}}
\def\tq{{\tilde{Q}}}

\def\tu{{\tilde{u^c}}}

\def\unity{{\hbox{1\kern-.6mm l}}}
\newcommand{\ov}{\overline}

\newcommand{\no}{\nonumber}

\newcommand{\real}{{\rm Re}\,}
\newcommand{\BR}{{\rm BR}}
\newcommand{\CR}{{\rm CR}}

\newcommand{\gev}{\mbox{GeV}}
\newcommand{\tev}{\mbox{TeV}}
\newcommand{\ga}{\gamma}

\newcommand{\gsim}{\lower.7ex\hbox{$\;\stackrel{\textstyle>}{\sim}\;$}}
\newcommand{\lsim}{\lower.7ex\hbox{$\;\stackrel{\textstyle<}{\sim}\;$}}

\newcommand{\captions}{\sf \caption}

\begin{document}

\begin{titlepage}

\begin{flushright}
\vglue -1.5cm
{CERN-PH-TH/2010-157}\\
{DFPD-10/TH/07}
\end{flushright}

\vglue .5cm
\begin{center}

{\LARGE \bf
Beyond the standard seesaw:
\\
\vspace{0.2cm} neutrino masses from K\"ahler operators
\\
\vspace{0.4cm} and broken supersymmetry }

\vspace{1cm}

{\large
Andrea Brignole$^{a}$, Filipe R.
Joaquim$^{b,\star}$ and Anna Rossi$^{c}$}
\\[7mm]
{\it $^a$ INFN, Sezione di Padova, I-35131 Padua, Italy \\[1.5mm]
\it $^b$ CERN, Theory Division, CH-1211 Geneva 23, Switzerland}\\[1.5mm]
{\it $^c$ Dipartimento di Fisica ``G.~Galilei'', Universit\`a di
Padova, I-35131 Padua, Italy
}

\vspace{1.0cm}

\begin{abstract}

\baselineskip=18pt

We investigate supersymmetric scenarios in which neutrino masses are
generated by effective $d=6$ operators in the K\"ahler potential,
rather than by the standard $d=5$ superpotential operator. First, we
discuss some general features of such effective operators, also
including SUSY-breaking insertions, and compute the relevant
renormalization group equations. Contributions to neutrino masses
arise at low energy both at the tree level and through finite
threshold corrections. In the second part we present simple explicit
realizations in which those K\"ahler operators arise by integrating
out heavy $SU(2)_W$ triplets, as in the type II seesaw. Distinct
scenarios emerge, depending on the mechanism and the scale of
SUSY-breaking mediation. In particular, we propose an appealing and
economical picture in which the heavy seesaw mediators are also
messengers of SUSY breaking. In this case, strong correlations exist
among neutrino parameters, sparticle and Higgs masses, as well as
lepton flavour violating processes. Hence, this scenario can be
tested at high-energy colliders, such as the LHC, and at lower
energy experiments that measure neutrino parameters or search for
rare lepton decays.

\end{abstract}

\end{center}

\vspace*{1.5cm}
\hspace*{0.2cm}\rule[0cm]{10cm}{0.01cm} \\
{\footnotesize \hspace*{0.8cm}$^\star${\rm On leave from ``Centro de
F\'{i}sica Te\'orica de Part\'{i}culas (CFTP)'', Lisbon,
Portugal.}\\
\tt \hspace*{0.8cm}\,E-mail addresses:
\hspace*{-0.1cm}brignole@pd.infn.it,
\hspace*{-0.1cm}filipe.joaquim@cern.ch,
\hspace*{-0.1cm}arossi@pd.infn.it }

\end{titlepage}


\baselineskip=18pt

\section{Introduction}

The seesaw mechanism can be regarded as a paradigm to explain the
smallness of neutrino masses. In the simplest scenarios, neutrinos
acquire Majorana masses scaling as $m_\nu \sim v^2 /M$, where $v$ is
the electroweak scale and $M \gg v$ is a heavy mass. The
experimental neutrino data \cite{nuexp,nupheno} point towards a
natural value $M \sim  10^{15}  \,\gev$, close to the Grand
Unification scale. From a low-energy perspective, the $1/M$
dependence appears as the coefficient of the lowest dimension
($d=5$) $SU(2)_W \times U(1)_Y$ invariant operator which violates
lepton number by two units ($\Delta L =2$), namely $(H L)^2/M$,
where $H$ and $L$ are Higgs and lepton doublets \cite{wein}. From a
more fundamental perspective, this effective operator usually arises
from integrating out heavy states with mass $\sim M$. At the tree
level, such heavy seesaw mediators can be either singlet `neutrinos'
coupled to $H L $ (type I \cite{type1}), $SU(2)_W$ triplet scalars
with non-zero hypercharge coupled to $L L$ and $H H$ (type II
\cite{type2}) or $SU(2)_W$ triplet fermions with zero hypercharge
coupled to $H L$ (type III \cite{type3}). These realizations can be
also implemented in supersymmetric (SUSY) extensions of the Standard
Model (SM). In such models, which contain two Higgs
superfields\footnote{Following the standard notation, we will use
the same symbol for a Higgs (matter) chiral superfield and its
scalar (fermionic) component field.} $H_1$ and $H_2$ with opposite
hypercharges, the leading $\Delta L =2$ effective operator is the
$d=5$ superpotential operator $\int \! d^2 \theta \, (H_2 L)^2/M$.

The scaling of neutrino masses with $v^2 /M$ is not the only
possibility, though. It is also conceivable that neutrino masses are
suppressed by a higher power of the heavy scale $M$, the simplest
possibility beyond $1/M$ being
\be\label{d6}
m_\nu \sim \frac{m\, v^2}{M^2}  \; ,
\ee
where $m\ll M$ is another mass parameter. In fact, the non-SUSY type
II seesaw \cite{type2} generically leads to neutrino masses
depending on two mass parameters, like in eq.\,(\ref{d6}). The same
occurs in suitable variants of the type I seesaw (see \eg \,
\cite{valle}) or in some radiative mechanisms \cite{zee}. In other
cases, $m_\nu$ is suppressed by even higher powers of $M$ (see \eg
\, \cite{high}).

In this work, we will focus on SUSY models where neutrino masses
behave like in eq.\,(\ref{d6}), with the additional requirement that
$m$ is related to the electroweak scale. Consequently, $M$ is
naturally lowered to intermediate values $M \lsim 10^{9}\,\gev$. The
behaviour described by eq.\,(\ref{d6}) can be realized in various
ways, through either K\"{a}hler or superpotential $d=6$, $\Delta L
=2$ effective operators. In the latter case, a possible operator is
$\int \! d^2 \theta \, S(H_2 L)^2/M^2$, where $S$ is a SM singlet
with $\langle S \rangle \sim v$. This fits naturally in the
framework of the next-to-minimal SUSY SM (NMSSM)~\cite{nmssm}, for
which some tree-level realizations have been proposed in \cite{GOS}.
Regarding $d=6$ K\"{a}hler operators, two candidates have been
pointed out in \cite{CEN}, namely $\int  \! d^4 \theta \,
(H^\dagger_1L) (H_2 L)/M^2$ and $\int \!  d^4 \theta \, (H^\dagger_1
L)^2/M^2$. In this case, neutrino masses arise in the form of
(\ref{d6}) with $m \sim \mu$, where the superpotential parameter
$\mu$ emerges from the replacement $F^\dagger_{H_1} \rightarrow -\mu
H_2$.

The purpose of this article is to generalize the proposal of
\cite{CEN} in several directions.  In Section~2 we describe some
general features of $d=6$, $\Delta L =2$ effective operators and
then focus on the K\"{a}hler operator $(H^\dagger_1 L)^2$. We point
out the importance of including SUSY-breaking insertions and find
novel contributions to neutrino masses of the form (\ref{d6}), in
which $m$ is a SUSY-breaking mass. Such SUSY-breaking contributions
can be even the dominant source of neutrino masses. We also discuss
and evaluate two classes of quantum effects related to those
operators: low-energy finite corrections at the sparticle threshold
and logarithmic corrections above it, described by renormalization
group equations (RGEs). In Section~3 we move from the effective
level to a more fundamental one and present the simplest explicit
realization of the K\"{a}hler operator $(H^\dagger_1 L)^2$,
including SUSY-breaking effects. This ultra-violet (UV) completion
of the effective theory is obtained in a type II seesaw framework.
The SUSY-breaking parameters associated with $(H^\dagger_1 L)^2$ are
related to those of the heavy triplet states in the case
SUSY-breaking mediation occurs at or above the triplet scale,
otherwise they can be generated radiatively, \eg \, by low-scale
gauge mediation and RGEs.

Another interesting feature of the SUSY type II seesaw is that it
provides the simplest realization of minimal lepton flavour
violation (LFV), in the sense that the high and low-energy flavour
structures are directly related \cite{AR}. In Section~4 we present
an appealing and predictive version of the type II seesaw where the
heavy triplets, which generate the $\Delta L =2$ effective operators
at the tree level, are identified with the SUSY-breaking mediators,
responsible for generating sparticle masses at the quantum level
through gauge and Yukawa interactions. This scenario is a variant of
that proposed in \cite{JR} and relates neutrino and sparticle masses
even more closely, since their common source is the SUSY-breaking
holomorphic mass term of the heavy states. In particular, we compute
the full set of MSSM SUSY-breaking terms at the heavy triplet mass
scale (Section~4.1) and obtain the tree-level and quantum
contributions to the neutrino mass matrix (Section~4.2). We also
discuss the phenomenological viability of this scenario, the general
properties of the MSSM spectrum and the prospects for searches at
the CERN Large Hadron Collider (LHC) (Section~4.3). Special emphasis
is devoted to LFV effects, which distinguish our model from purely
gauge-mediated ones. In particular, we discuss the predictions for
charged-lepton radiative decays $\ell_i\rightarrow \ell_j\gamma$,
taking into account the near-future experimental sensitivity for
both such LFV searches and the measurements of neutrino parameters
(Section~4.4). Finally,
in Section~5 we summarise our results and draw the concluding remarks.\\

\section{Neutrino masses from K\"{a}hler operators}

As anticipated in the Introduction, we are interested in SUSY
scenarios in which neutrino masses are generated by effective $d=6$,
$\Delta L = 2$ operators. Before focussing on a specific class  of
such operators, let us briefly describe some of their general
features.

\subsection{Lepton number violating $d=6$ operators}

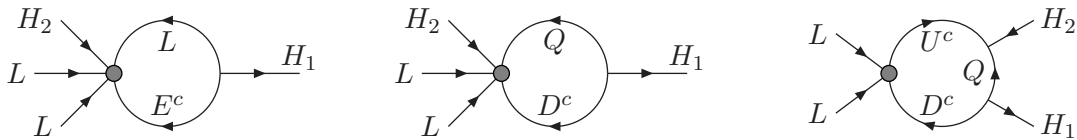
\begin{figure}[b]
\vspace{1.0 cm}
\begin{center}
\begin{picture}(100,50)(-50,-25)
\ArrowArc(0,0)(20,0,180)
\ArrowArcn(0,0)(20,0,180)
\ArrowLine(-50,0)(-20,0)
\ArrowLine(-40,20)(-20,0)
\ArrowLine(-40,-20)(-20,0)
\ArrowLine(20,0)(50,0)
\GCirc(-20,0){3}{0.5}
\Text(-57,0)[]{\small $L$}
\Text(-47,-20)[]{\small $L$}
\Text(-50,20)[]{\small $H_2$}
\Text(50,6)[]{\small $H_1$}
\Text(0,12)[]{\small $L$}
\Text(0,-12)[]{\small $E^c$}
\end{picture}
\hglue 1.5cm
\begin{picture}(100,50)(-50,-25)
\ArrowArc(0,0)(20,0,180)
\ArrowArcn(0,0)(20,0,180)
\ArrowLine(-50,0)(-20,0)
\ArrowLine(-40,20)(-20,0)
\ArrowLine(-40,-20)(-20,0)
\ArrowLine(20,0)(50,0)
\GCirc(-20,0){3}{0.5}
\Text(-57,0)[]{\small $L$}
\Text(-47,-20)[]{\small $L$}
\Text(-50,20)[]{\small $H_2$}
\Text(50,6)[]{\small $H_1$}
\Text(0,12)[]{\small $Q$}
\Text(0,-12)[]{\small $D^c$}
\end{picture}
\hglue 1.5cm
\begin{picture}(100,50)(-50,-25)
\ArrowArcn(0,0)(20,180,30)
\ArrowArcn(0,0)(20,-30,180)
\ArrowArc(0,0)(20,-30,30)
\ArrowLine(-40,15)(-20,0)
\ArrowLine(-40,-15)(-20,0)
\ArrowLine(34.6,20)(17.3,10)
\ArrowLine(17.3,-10)(34.6,-20)
\GCirc(-20,0){3}{0.5}
\Text(-47,15)[]{\small $L$}
\Text(-47,-15)[]{\small $L$}
\Text(44,20)[]{\small $H_2$}
\Text(44,-20)[]{\small $H_1$}
\Text(-2,12)[]{\small $U^c$}
\Text(-2,-12)[]{\small $D^c$}
\Text(12,0)[]{\small $Q$}
\end{picture}
\end{center}
\caption{\small Logarithmically divergent contributions to
the  K\"ahler operator $(H^\dagger_1 L) (H_2 L)/M^2$
induced by other $d=6$ operators.}
\label{f1}
\end{figure}

Consider an effective low-energy theory with the field content of
the minimal SUSY SM (MSSM)~\cite{mssm} and conserved R-parity. The
leading $\Delta L = 2$ operator is the well-known $d=5$ term $(H_2
L)^2/M \subset W$. In case this operator is (for some reason)
suppressed or absent, there are two $d=6$, $\Delta L = 2$ K\"ahler
operators which can generate neutrino masses, namely $(H^\dagger_1L)
(H_2 L)/M^2 \subset K$ and $(H^\dagger_1 L)^2/M^2 \subset
K$~\cite{CEN}. We remark that, in principle, additional $d=6$,
$\Delta L = 2$ operators of the form $ LLLE^c H_2/M^2 \subset W$,
$LLQ D^c H_2/M^2 \subset W$ and $ LL {U^c}^\dagger \! D^c/M^2\subset
K$ should be considered as well. Indeed, although these do not
generate neutrino masses at the tree level, they do so radiatively,
by inducing $(H^\dagger_1L) (H_2 L)/M^2 \subset K$ via RGEs (see
Fig.~\ref{f1}).
%
%
%
The fact that some operators mix under renormalization suggests that
all the above $\Delta L = 2$ operators could be grouped in distinct
classes, by means of continuous or discrete Peccei-Quinn (PQ)-like
symmetries under which $H^\dagger_1$ and $H_2$ transform differently
and the ordinary Yukawa couplings are invariant. Since $\mu H_2 H_1
\subset W$ breaks such symmetries, we can consider the small $\mu$
parameter ($\mu \ll M$) as a `minimal' effective PQ spurion, such
that non-invariant operators in $K$ and $W$ will be suppressed by
powers of $\epsilon =\mu/\Lambda_*$, where $\Lambda_*\geq M$ is some
high scale (\eg, that where $\mu$ is generated). For instance, if
$(H^\dagger_1 L)^2/M^2 \subset K$ is allowed, we expect the
remaining operators to be subleading, since the set
$\{(H^\dagger_1L) (H_2 L)/M^2, LLLE^c H_2/M^2, LLQ D^c H_2/M^2, LL
{U^c}^\dagger \! D^c/M^2 \}$ will be suppressed by a factor
$\epsilon$ while $(H_2 L)^2/M$ will be suppressed by $\epsilon^2$.
Alternatively, the symmetries may allow the operators $\{
(H^\dagger_1L) (H_2 L)/M^2, LLLE^c H_2/M^2, LLQ D^c H_2/M^2, LL
{U^c}^\dagger \! D^c/M^2 \}$ and suppress the others.

Similar arguments can be applied to extensions of the MSSM in which
$\mu$ is effectively generated at low energies by the vacuum
expectation value (VEV) of a SM singlet $S$, through the term $S H_2
H_1 \subset W$. The field $S$ is usually charged under some
symmetry, such as a $Z_3$ in the NMSSM or an extra gauged $U(1)$.
Such symmetries provide selection rules for the effective operators
as well. For instance, in the NMSSM framework $Z_3$ selection rules
were used in \cite{GOS} to generate neutrino masses at leading order
through either $(H_2 L)^2/M \subset W$ ($d=5$), $S(H_2 L)^2/M^2
\subset W$ ($d=6$) or $S^2(H_2 L)^2/M^3 \subset W$ ($d=7$). Notice
that only one of these operators can be invariant since each of them
carries a different $Z_3$ charge. Nevertheless, if the allowed
operator of this list has $d>5$, other $\Delta L = 2$ operators are
also allowed in $W$ or $K$ with the same or lower dimensionality,
which were not considered in \cite{GOS}. As an example, suppose
$S(H_2 L)^2/M^2 \subset W$ is allowed by $Z_3$. Then, also the $d=6$
set $\{(H^\dagger_1L) (H_2 L)/M^2, LLLE^c H_2/M^2, LLQ D^c H_2/M^2,
LL {U^c}^\dagger \! D^c/M^2 \}$ is allowed. Alternatively, if $S^2
(H_2 L)^2/M^3 \subset W $ is $Z_3$ symmetric, then several other
operators of the same dimension ($d=7$) such as $S(H^\dagger_1L)
(H_2 L)/M^3$, $S^\dagger (H_2 L)^2/M^3$, $S LL {U^c}^\dagger \!
D^c/M^2 \subset K$ or $S LLLE^c H_2/M^3$, $S LLQ D^c H_2/M^3$, $H_1
H_2 (H_2 L)^2/M^3 \subset W$ are permitted. Even more importantly,
in this case there is a single $Z_3$-invariant operator of lower
dimension, namely the $d=6$ term $(H^\dagger_1 L)^2/M^2 \subset K$.

The above discussion (which extends those of \cite{CEN,GOS})
emphasizes the fact that symmetry arguments in the effective theory
can partly justify the assumption that a specific operator dominates
over others. The ultimate motivation for such a selection should lie
at a more fundamental level, \ie, in the UV completion of the
effective theory. From a minimal low-energy perspective, we note
that the $d=6$ term $(H^\dagger_1 L)^2/M^2 \subset K$ is somehow
singled out in the above examples by its symmetry properties.
Furthermore, in Section~3 we will show that this operator admits a
very simple tree-level realization.


\subsection{K\"ahler operators $(H^\dagger_1 L)^2/M^2$ with broken SUSY}

We proceed with our discussion by assuming that the leading $\Delta
L=2$ effective operator has the form $(H^\dagger_1 L)^2/M^2$. In
general, we expect it to be accompanied by analogous operators with
SUSY-breaking insertions\footnote{ Effective operators with $d>4$
and SUSY-breaking insertions have been also considered in other
contexts, such as the Higgs sector \cite{higgs} or baryon number
violation \cite{proton}. SUSY-breaking effects in the neutrino
sector have been considered in \cite{susynu} from a perspective
which is different from ours. In those works, SUSY breaking was
invoked to suppress either $L H_2 N$ Yukawa couplings or $M_N N N$
mass terms (or both) in models with singlet states $N$.} of the form
$X/M_S$, ${X^\dagger}/{M_S}$, $ X X^\dagger/{M_S^2}$, where
$X=\theta^2 F_X$ is a SUSY-breaking spurion superfield (VEVs are
understood) and $M_S$ is the scale of SUSY-breaking mediation, which
could be either larger or smaller than $M$. It is also tempting to
identify $M_S$ with $M$, as we will do in Section~4.

In general, we can write the relevant $\Delta L=2$ effective
lagrangian as
\be
\label{K1}
 {\cal L}_{\rm eff}  =
\int \!   d^4 \theta \frac{1}{2 M^2} \left( \bkappa
+ \bbe_\kappa \frac{X}{M_S}
+ \tilde{\bbe}_\kappa \frac{X^\dagger}{M_S}
+ \bg_\kappa \frac{X X^\dagger}{M_S^2}
\right)_{ij} (H^\dagger_1 L_i) (H^\dagger_1 L_j)
\,\, + {\rm h.c.}\,,
\ee
where $i,j=e,\mu,\tau$ are flavour indices, $\bkappa, \bbe_\kappa,
\tilde{\bbe}_\kappa, \bg_\kappa$ are dimensionless flavour-dependent
parameters and the SUSY gauge completion $(H^\dagger_1 L)
\rightarrow (H^\dagger_1 e^{2 V} L)$ is understood. In principle, we
could have incorporated the factor $1/M^2$ into dimensionful
coefficients as, for instance, $\bkappa'=\bkappa/M^2$. This would
better suit models in which the masses of the heavy states to be
integrated out carry a flavour structure. However, even in such
cases one can always factor out an overall $1/M^2$. We have chosen
the parametrization (\ref{K1}) to exhibit mass dimensions in a more
transparent way, and also because the explicit realizations
presented in Sections~3 and~4 make use of heavy states with
unflavoured masses.

By replacing $X$ with its SUSY-breaking VEV in eq.~(\ref{K1}), we
obtain the equivalent parametrization
\be
\label{K2}
{\cal L}_{\rm eff}  =
\int  \!  d^4 \theta \frac{1}{2 M^2} \left( \bkappa + \theta^2 \bB_\kappa +
\bar{\theta}^2 \tilde{\bB}_\kappa +  \theta^2  \bar{\theta}^2 \bC_\kappa
\right)_{ij} (H^\dagger_1 L_i) (H^\dagger_1 L_j)
\,\, + {\rm h.c.}\,,
\ee
where we have traded the coefficients $\bbe_\kappa,
\tilde{\bbe}_\kappa$ and $\bg_\kappa$ for dimensionful SUSY-breaking
parameters $ \bB_\kappa =  \bbe_\kappa F_X/M_S$, $\tilde{\bB}_\kappa
= \tilde{\bbe}_\kappa F^*_X / M_S$ (both of dimension one) and
$\bC_\kappa =  \bg_\kappa |F_X|^2/M_S^2$ (of dimension two),
respectively. The magnitude and flavour structure of all these
parameters depend on the underlying physics which generates them. In
Sections~3 and 4 we will show explicit realizations which lead to
simple correlations among the above quantities. Here, we will keep
our discussion at a general and model-independent level. Notice that
the SUSY part of ${\cal L}_{\rm eff}$ is generated at $M$, while the
SUSY-breaking one emerges at scales below $\min (M,M_S)$. At low
energy, all four operators in eq.~(\ref{K1}) [or, equivalently, in
eq.~(\ref{K2})] contribute to neutrino masses either directly or
indirectly, as we will show later. Before doing that, we will
discuss the connection between high and low energies, namely the
renormalization group evolution of the effective operators.


\subsection{Renormalization group evolution}

A convenient tool to derive the RGEs for the $\Delta L=2$ operators shown in
eq.~(\ref{K2}) is the general
expression of the one-loop corrected  K\"ahler potential obtained in
\cite{ab}, which applies to general effective SUSY theories with
K\"ahler potential $K(\phi,\phi^*)$, superpotential $W(\phi)$ and
gauge kinetic function $f_{ab}(\phi)$. The logarithmically divergent
correction to $K$ reads \cite{ab}:
\be
\label{deltak} (\Delta K)_{\log} = \frac{\log \Lambda_{\rm
UV}^{\!2}}{32 \pi^2} \left[ W_{ij} K^{j \bar{m}} W^*_{\bar{m}
\bar{n}} K^{i \bar{n}} - 4 \, (\real f_a)^{-1} \, ( \phi^\dagger
T^a)^{\bar{\imath}} K_{\bar{\imath} j} (T^a \phi)^j \right]\,,
\ee
where $\Lambda_{\rm UV}$ is an UV cutoff,
 $W_{ij}=\partial^2 W/\partial \phi_i
\partial \phi_j$, $K_{\bar{\imath} j} = \partial^2 K / \partial
\phi^*_{\bar{\imath}}
\partial \phi_j $, $K^{i \bar{n}} K_{\bar{n} j} = \delta^i_j$,
$T^a$ are the generators of the gauge group, and we have considered
a diagonal kinetic function $f_{ab} = f_a \delta_{ab}$. By applying
eq.~(\ref{deltak}) to our case, one can extract the corrections to
the K\"ahler terms $H_1^\dagger H_1$, $L^\dagger L$ and
$(H_1^\dagger L)^2$. The relevant RGEs are derived by combining wave
function and vertex corrections. In fact, the RGE for $\bkappa$ was
obtained in this way in \cite{CEN}. This method allows us to derive
the RGEs for $\bB_\kappa$, ${\tilde \bB}_\kappa$ and $\bC_\kappa$ as
well, by retaining the dependence of $K$, $W$ and $f_a$ on the
spurion superfield $X= \theta^2 F_X$, which effectively generates
all SUSY-breaking mass parameters. We recall that gaugino masses
appear in $f_a=\frac{1}{g_a^2}(1- 2 \theta^2 M_a)$, scalar masses
stem from $K$ as $(1-\theta^2 \bar{\theta}^2
\tilde{m}^2)\phi^\dagger \phi$, while Yukawa and SUSY-breaking
trilinear couplings come from $W$ through combinations like $(\bY_e
- \theta^2 \bA_e) H_1 E^c L$. Some of the loop-induced terms in
$(\Delta K)_{\log}$ have the form $\theta^2 \phi^\dagger \phi$ (or
$\bar{\theta}^2 \phi^\dagger \phi$) and can be included in a
$\theta$-dependent ($\bar{\theta}$-dependent) wave function
renormalization of the superfield $\phi$
($\phi^\dagger$)~\cite{yam}. Our final result for the RGEs is:
\bea
\label{rgek}
8 \pi^2  \frac{d \bkappa}{dt } & = &
 \left[ g^2 + g'^2
+  {\rm Tr}( \bY_e^\dagger \bY_e + 3 \bY^\dagger_d  \bY_d)
\right]\bkappa -  \frac{1}{2}\left[ \bkappa \bY^\dagger_e \bY_e +
(\bY^\dagger_e \bY_e )^T  \bkappa \right]\,,
\\
\label{rgebk}
8 \pi^2  \frac{d \bB_\kappa}{dt } & = & \left[ g^2 +
g'^2 +  {\rm Tr}( \bY_e^\dagger \bY_e + 3 \bY^\dagger_d  \bY_d)
\right]  \bB_\kappa - \frac{1}{2}\left[ \bB_\kappa \bY^\dagger_e \bY_e +
 (\bY^\dagger_e \bY_e )^T  \bB_\kappa \right]
\nonumber \\
& & \!\!\! + \left[ g^2 M_2 + g'^2 M_1 \right] \bkappa\,,
\\
\label{rgebkt}
8 \pi^2  \frac{d {\tilde \bB}_\kappa}{dt } & = & \left[
g^2 + g'^2 +  {\rm Tr}( \bY_e^\dagger \bY_e + 3 \bY^\dagger_d
\bY_d)  \right] {\tilde \bB}_\kappa   -  \frac{1}{2}\left[
{\tilde \bB}_\kappa \bY^\dagger_e \bY_e + (\bY^\dagger_e \bY_e )^T
{\tilde \bB}_\kappa \right]
\nonumber \\
& &  \!\!\! + \left[ g^2 M^*_2 + g'^2 M^*_1 - 2 \, {\rm Tr}(
\bA_e^\dagger \bY_e + 3 \bA^\dagger_d  \bY_d)  \right] \bkappa +
\bkappa \, \bA^\dagger_e \bY_e + (\bA^\dagger_e \bY_e )^T \bkappa\,,
\\
\label{rgeck}
8 \pi^2  \frac{d \bC_\kappa}{dt } & = & \left[ g^2 +
g'^2 +  {\rm Tr}( \bY_e^\dagger \bY_e + 3 \bY^\dagger_d  \bY_d)
\right]  \bC_\kappa - \frac{1}{2} \left[ \bC_\kappa \bY^\dagger_e \bY_e
+ (\bY^\dagger_e \bY_e )^T \bC_\kappa \right]
\nonumber \\
& &  \!\!\! + \left[ g^2 M^*_2 + g'^2 M^*_1 - 2 \, {\rm Tr}(
\bA_e^\dagger \bY_e + 3 \bA^\dagger_d  \bY_d)  \right]  \bB_\kappa +
\bB_\kappa \bA^\dagger_e \bY_e + (\bA^\dagger_e \bY_e )^T  \bB_\kappa
\nonumber \\
& &  \!\!\! + \left[ g^2 M_2 + g'^2 M_1 \right] {\tilde \bB}_\kappa
+ 4 \, [\, 2 \, g^2 |M_2|^2 + g'^2 |M_1|^2 \,] \bkappa - \bkappa \,
\bP - \bP^T \bkappa\,,
\eea
where $ \bP \equiv \bA^\dagger_e \bA_e + (\bmm^2_{\tl})^T
\bY_e^\dagger \bY_e + \bY_e^\dagger (\bmm^2_{\te})^T \bY_e +
m^2_{H_1} \bY_e^\dagger \bY_e$. These RGEs hold in the MSSM or in
its extensions with extra states that do not couple to either $H_1$
or $L$. The generalization to models with such extra couplings is
straightforward. For instance, the NMSSM superpotential couplings
$(\la_S -\theta^2 A_S) S H_2 H_1$ only lead to a few extra terms in
the RGEs. In practice, it is enough to shift $ {\rm Tr}(
\bY_e^\dagger \bY_e) \rightarrow
 {\rm Tr}( \bY_e^\dagger \bY_e) + |\la_S|^2 $ and
${\rm Tr}(\bA_e^\dagger \bY_e)  \rightarrow
{\rm Tr}(\bA_e^\dagger \bY_e) + A_S^* \la_S$ in the above equations.

In general, eqs.~(\ref{rgek})-(\ref{rgeck}) form a system of coupled
RGEs, which exhibits operator mixing. Each equation contains a
`homogeneous' part, which is common to all four operators. Those
 of ${\bB}_\kappa$ and ${\tilde \bB}_\kappa $
have an additional piece which is driven by $\bkappa$ and depends on
the gaugino masses and trilinear couplings. As for the RGE of
$\bC_\kappa$, its inhomogenous part also contains ${\bB}_\kappa$ and
${\tilde \bB}_\kappa $. Notice that the RGEs involve several
independent parameters and flavour structures. Still, important
simplifications may occur in specific scenarios (see Sections~3 and
4).

For completeness, we also present the RGEs of the
$d=5$ superpotential operator $\int \!
d^2 \theta \frac{1}{2 M_5}(\bkappa_5 + \theta^2 \bB_5)_{ij}(L_i
H_2)(L_j H_2)$:
\bea
\label{rgeh} 8 \pi^2  \frac{d \bkappa_5}{dt } & = & - \left[ 3 g^2 +
g'^2 - 3  {\rm Tr}( \bY_u^\dagger \bY_u) \right] \bkappa_5 +
\frac{1}{2}\left[ \bkappa_5 \bY^\dagger_e \bY_e + (\bY^\dagger_e
\bY_e )^T \bkappa_5 \right]\,,
\\
 \label{rgebh} 8 \pi^2  \frac{d \bB_5}{dt } & = & - \left[
3 g^2 + g'^2 -3 {\rm Tr}( \bY_u^\dagger \bY_u )  \right]  \bB_5 +
\frac{1}{2}\left[ \bB_5 \bY^\dagger_e \bY_e + + (\bY^\dagger_e \bY_e
)^T \bB_5 \right]
\nonumber \\
& &  \!\!\! -2 \left[ 3 g^2 M_2 + g'^2 M_1 + 3 \, {\rm Tr}(
\bY_u^\dagger \bA_u) \right] \bkappa_5 - \bkappa_5 (\bY^\dagger_e
\bA_e) - (\bY^\dagger_e \bA_e )^T \bkappa_5\,.
\eea
The equation for $\bkappa_5$ is well known \cite{rged5}, while that
for the corresponding SUSY-breaking parameter $\bB_5$ is another
novel result.


\subsection{Tree-level contributions to neutrino masses}

\begin{figure}[]
\vspace{0.4cm}
\begin{center}
\begin{picture}(120,80)(-60,-40)
\ArrowLine(-50,0)(0,0)
\ArrowLine(50,0)(0,0)
\DashArrowLine(0,0)(-40,40){3}
\DashArrowLine(0,0)(20,20){3}
\DashArrowLine(40,40)(20,20){3}
\GCirc(0,0){3}{0.5}
\Text(-50,-8)[]{\small $L$}
\Text(50,-8)[]{\small $L$}
\Text(-25,40)[]{\small $H_1$}
\Text(4,19)[]{\small $F_{H_1}$}
\Text(25,40)[]{\small $H_2$}
\Text(20,20)[]{$+$}
\Text(27,17)[]{\small $\mu$}
\Text(0,-12)[]{$\frac{\kappa}{M^2}$}
\end{picture}
\hglue 2.5cm
\begin{picture}(120,80)(-60,-40)
\ArrowLine(-50,0)(0,0)
\ArrowLine(50,0)(0,0)
\DashArrowLine(0,0)(-40,40){3}
\DashArrowLine(0,0)(40,40){3}
\GCirc(0,0){3}{0.5}
\Text(-50,-8)[]{\small $L$}
\Text(50,-8)[]{\small $L$}
\Text(-25,40)[]{\small $H_1$}
\Text(25,40)[]{\small $H_1$}
\Text(1,-13)[]{$\frac{\tilde{B}_\kappa}{M^2}$}
\end{picture}
\end{center}
\vspace{-1.0 cm}
\caption{\small Tree-level contributions to neutrino masses from the
K\"ahler operators $(H^\dagger_1 L)^2$ and $X^\dagger (H^\dagger_1 L)^2$.}
\label{f2}
\end{figure}
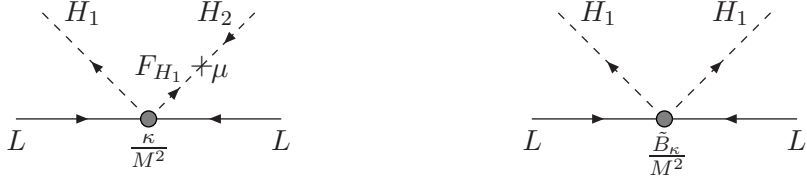

Consider now the effective K\"ahler operators of eq.~(\ref{K2}),
renormalized at the weak scale. Two of them, namely those with
coefficients $\bkappa$ and $\tilde{\bB}_\kappa$ (see Fig.~\ref{f2}),
contribute directly to the neutrino mass matrix (${\cal L} \supset
-\frac{1}{2}(\bmm_\nu)_{ij} \nu_i \nu_j + {\rm h.c.}$):
\be
\label{mnu} \bmm_\nu = \bmm^{(\kappa)}_\nu +
\bmm^{(\tilde{B}_\kappa)}_\nu\,.
\ee
The $\bkappa$-operator gives a lagrangian operator of the form $
(F_{H_1}^\dagger L)(H_1^\dagger L)$, which reduces to $ -\mu (H_2
L)(H_1^\dagger L)$ after replacing $F_{H_1}^\dagger \rightarrow -\mu
H_2$. Upon setting the Higgs fields to their VEVs ($\langle H_1^0
\rangle = v \cos\beta$, $\langle H_2^0 \rangle = v \sin\beta$) one
gets \cite{CEN}
\be
\label{mnuk} \bmm^{(\kappa)}_\nu = 2\,  \bkappa \;  \mu \, \frac{
v^2}{M^2} \sin\beta \cos\beta\,.
\ee
On the other hand, the $\tilde{\bB}_\kappa$-operator leads to a
lagrangian term of the form $ (H_1^\dagger L)^2$, which induces
\be
\label{mnubtk} \bmm^{(\tilde{B}_\kappa)}_\nu = \tilde{\bB}_\kappa
\frac{v^2}{M^2} \cos^2\beta\,.
\ee
This novel contribution to the neutrino mass matrix can be of the
same order of $\bmm^{(\kappa)}_\nu$. In general, both the flavour
structure and the relative size of $\bmm^{(\kappa)}_\nu$ and
$\bmm^{(\tilde{B}_\kappa)}_\nu$ are model dependent. For instance, a
large value of $\tan\beta$ suppresses
$\bmm^{(\tilde{B}_\kappa)}_\nu$ with respect to
$\bmm^{(\kappa)}_\nu$, whereas a hierarchy $\tilde{\bB}_\kappa \gg
\bkappa \mu$ enhances it (see Sections~3 and 4).

\subsection{Radiative finite contributions to $\bmm_\nu$}

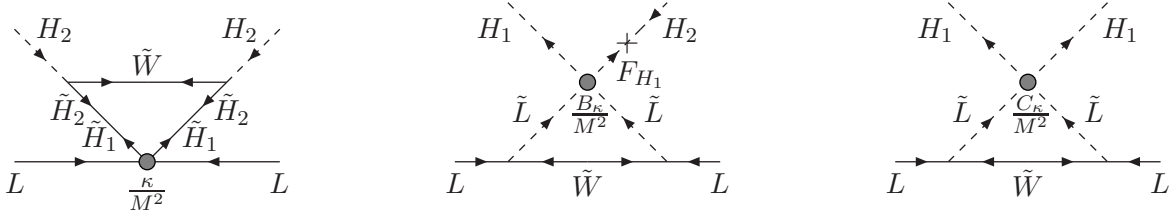
\begin{figure}[]
\vspace{1.0 cm}
\begin{center}
\begin{picture}(120,80)(-60,-40)
\ArrowLine(-50,0)(0,0) \ArrowLine(50,0)(0,0) \ArrowLine(0,0)(-15,15)
\ArrowLine(-30,30)(-15,15) \DashArrowLine(-50,50)(-30,30){3}
\ArrowLine(0,0)(15,15) \ArrowLine(30,30)(15,15)
\DashArrowLine(50,50)(30,30){3} \ArrowLine(-30,30)(0,30)
\ArrowLine(30,30)(0,30) \GCirc(0,0){3}{0.5} \Text(-50,-8)[]{\small
$L$} \Text(50,-8)[]{\small $L$} \Text(-18,10)[]{\small ${\tilde
H}_1$} \Text(-30,20)[]{\small ${\tilde H}_2$} \Text(20,10)[]{\small
${\tilde H}_1$} \Text(32,20)[]{\small ${\tilde H}_2$}
\Text(0,38)[]{\small ${\tilde W}$} \Text(-35,50)[]{\small $H_2$}
\Text(35,50)[]{\small $H_2$} \Text(0,-12)[]{$\frac{\kappa}{M^2}$}
\end{picture}
\hglue 1.5cm
\begin{picture}(120,80)(-60,-40)
\ArrowLine(-50,0)(-30,0) \ArrowLine(50,0)(30,0)
\ArrowLine(0,0)(-30,0) \ArrowLine(0,0)(30,0)
\DashArrowLine(-30,0)(0,30){3} \DashArrowLine(30,0)(0,30){3}
\DashArrowLine(0,30)(-30,60){3} \DashArrowLine(0,30)(20,50){3}
\DashArrowLine(30,60)(20,50){3} \GCirc(0,30){3}{0.5}
\Text(-50,-8)[]{\small $L$} \Text(50,-8)[]{\small $L$}
\Text(-25,20)[]{\small ${\tilde L}$} \Text(25,20)[]{\small ${\tilde
L}$} \Text(-35,50)[]{\small $H_1$} \Text(20,33)[]{\small $F_{H_1}$}
\Text(35,50)[]{\small $H_2$} \Text(15,45)[]{$+$}
\Text(0,-8)[]{\small ${\tilde W}$}
\Text(1,18)[]{$\frac{B_\kappa}{M^2}$}
\end{picture}
\hglue 1.5cm
\begin{picture}(120,80)(-60,-40)
\ArrowLine(-50,0)(-30,0) \ArrowLine(50,0)(30,0)
\ArrowLine(0,0)(-30,0) \ArrowLine(0,0)(30,0)
\DashArrowLine(-30,0)(0,30){3} \DashArrowLine(30,0)(0,30){3}
\DashArrowLine(0,30)(-30,60){3} \DashArrowLine(0,30)(30,60){3}
\GCirc(0,30){3}{0.5} \Text(-50,-8)[]{\small $L$}
\Text(50,-8)[]{\small $L$} \Text(-25,20)[]{\small ${\tilde L}$}
\Text(25,20)[]{\small ${\tilde L}$} \Text(-35,50)[]{\small $H_1$}
\Text(35,50)[]{\small $H_1$} \Text(0,-8)[]{\small ${\tilde W}$}
\Text(1,18)[]{$\frac{C_\kappa}{M^2}$}
\end{picture}
\end{center}
\vspace{-1.0 cm} \caption{\small One-loop finite contributions to
neutrino masses from the K\"ahler operators $(H^\dagger_1 L)^2$, $X
(H^\dagger_1 L)^2$ and $X X^\dagger (H^\dagger_1 L)^2$ (from left to
right). These diagrams generate lagrangian operators of the form
$(H_2 L)^2$, $(H_2 L) (H^\dagger_1 L)$ and $(H^\dagger_1 L)^2$ (the
$W$-ino can be replaced by a $B$-ino everywhere). } \label{f3}
\end{figure}

Additional contributions to $\bmm_\nu$ arise from quantum effects.
Since we have already discussed the logarithmic renormalization
between high and low scales, we now turn to the analysis of finite
quantum corrections to $\bmm_\nu$ at the weak scale, \ie, at the
sparticle threshold (see Fig.~\ref{f3}). In particular, we will
focus on effects which exhibit potential enhancement factors, such
as large values of $\tan\beta$ or large mass ratios.

Consider again the $\bkappa$-operator of eq.~(\ref{K2}). One of its
lagrangian components is a four fermion operator with two leptons
and two higgsinos, which can be dressed by a finite Higgsino-gaugino
loop (left diagram in Fig.~\ref{f3}), generating an effective
lagrangian term of the form $(H_2 L)^2$. Curiously, this reminds the
`holomorphic' structure of the familiar $d=5$ superpotential
operator, but of course it cannot be interpreted that way (it arises
radiatively through SUSY breaking). The resulting contribution to
$\bmm_\nu$ is proportional to the tree-level term
$\bmm^{(\kappa)}_\nu$ [see eq.~(\ref{mnuk})]:
\be
\label{deltamk} \delta_\kappa \bmm_\nu = \frac{1}{64 \pi^2} \left(
\frac{g^2}{M_2^*} \, f_{\mu 2} + \frac{g'^2}{M_1^*} \, f_{\mu 1}
\right) \mu \, \tan\beta \,\, \bmm^{(\kappa)}_\nu\,,
\ee
where $f_{\mu a}= f(|\mu|^2/|M_a|^2)$ and
$f(x)=(x-1- \log x)/(x-1)^2$. Despite the potential
$\tan\beta$ enhancement, this correction is below $2 \%$
for any ratios $\mu/M_a$ and any $\tan\beta < 50$.

Concerning the $\bB_\kappa$ and $\bC_\kappa$-operators of
eq.~(\ref{K2}), the component expansion of the former includes a
lagrangian term $(F_{H_1}^\dagger \tilde{L})(H_1^\dagger \tilde{L})$
which gives  $ -\mu (H_2 \tilde{L})(H_1^\dagger \tilde{L})$, while
the latter leads to $(H_1^\dagger \tilde{L})^2$. Both these terms
generate small $\Delta L=2$ corrections to {\em sneutrino} masses
(${\cal L} \supset -\frac{1}{2} (\delta \bmm^2_{\tilde{\nu}})_{ij}
\tilde{\nu}_i \tilde{\nu}_j  + {\rm h.c.}$), namely
\be
\label{sneu}
\delta \bmm^2_{\tilde{\nu}} = - (2 \, \mu \, \bB_\kappa \sin\beta
\cos\beta + \bC_\kappa \cos^2  \! \beta) \frac{v^2}{M^2}  \; ,
\ee
which induce tiny splittings in the sneutrino spectrum. This
property of our $d=6$ operators generalizes a known effect of $d=5$
type I \cite{grhab} and type II \cite{JR} seesaw
realizations\footnote{In the $d=5$ case, the operators shown before
eqs.~(\ref{rgeh}) and (\ref{rgebh}) induce neutrino masses
$\bmm_\nu= \bkappa_5 \sin^2  \! \beta \, v^2/M_5$ as well as $\Delta
L=2$ sneutrino masses $ \delta \bmm^2_{\tilde{\nu}} = - (2 \, \mu^* \,
\bkappa_5 \sin\beta \cos\beta + \bB_5 \sin^2  \!
\beta) v^2/M_5$.
 }, and is
potentially relevant for the phenomenon of sneutrino oscillations~\cite{grhab}.
Furthermore, the presence of $\Delta L=2$ scalar operators
induces neutrino masses at the one-loop level, as in \cite{grhab,hkkflt}.
In our framework, the lagrangian terms
$(H_2 \tilde{L})(H_1^\dagger \tilde{L})$ and
$(H_1^\dagger \tilde{L})^2$  can be
dressed by finite slepton-gaugino loops (middle and right diagrams
in Fig.~\ref{f3}), inducing effective lagrangian operators of
the form $(H_2 L)(H^\dagger_1 L)$ and $(H^\dagger_1 L)^2$,
respectively.

%

In order to discuss the $\bB_\kappa$ and $\bC_\kappa$ contributions
to $\bmm_\nu$, we parametrize the soft mass matrix of `left-handed'
sleptons  $\tilde{L}$ as $\bmm^2_{\tl} = {\tilde m}_L^2 ( \unity +
\bD_L)$, where ${\tilde m}_L^2$ sets the overall mass scale and the
dimensionless matrix $\bD_L$ accounts for flavour
dependence\footnote{A flavour violating $\bD_L$ generically appears
in models in which neutrino masses arise through coupling to heavy
states. The first such examples in the type I \cite{borzmas} and
type II \cite{AR} seesaws relied on renormalization effects. In the
type II model presented in Section~4 a non-vanishing $\bD_L$ is
generated by finite radiative corrections at the scale of
SUSY-breaking mediation.}. At first order in $\bD_L$, the
contributions to the neutrino mass matrix induced by $\bB_\kappa$
and $\bC_\kappa$ are:
\bea
\label{deltambk}
\delta_{B_\kappa} \bmm_\nu  & \simeq &
\frac{1}{32 \pi^2}
\left[ -
\left( \frac{g^2}{M_2} \, f_{L2}
+\frac{g'^2}{M_1} \, f_{L1} \right) \bB_\kappa  \right.
\no \\
& & \left.
+
\left( \frac{g^2}{M_2} \, h_{L2}
+\frac{g'^2}{M_1} \, h_{L1} \right)
(\bB_\kappa  \, \bD_L + \bD_L^T \, \bB_\kappa )
\right] 2 \mu \frac{v^2}{M^2} \sin\beta \cos\beta
\\
\label{deltamck}
\delta_{C_\kappa} \bmm_\nu  & \simeq &
\frac{1}{32 \pi^2}
\left[ -
\left( \frac{g^2}{M_2} \, f_{L2}
+\frac{g'^2}{M_1} \, f_{L1} \right) \bC_\kappa  \right.
\no \\
& & \left. + \left( \frac{g^2}{M_2} \, h_{L2} +\frac{g'^2}{M_1} \,
h_{L1} \right) (\bC_\kappa  \, \bD_L + \bD_L^T \, \bC_\kappa )
\right] \frac{v^2}{M^2} \cos^2 \! \beta
\eea
where $f_{L a}= f ({\tilde m}_L^2/|M_a|^2)$, $h_{L a}= h({\tilde
m}_L^2/|M_a|^2)$, $h(x)=(x^2-1-2x \log x)/(x-1)^3$ and $f(x)$ was
defined after eq.~(\ref{deltamk}). Both the flavour structure and
the size of $\delta_{B_\kappa}\bmm_\nu$, $\delta_{C_\kappa}
\bmm_\nu$ are model dependent. The flavour dependence enters through
$\bB_\kappa$, $\bC_\kappa$ and $\bD_L$, while the overall size
crucially depends on the magnitude of the SUSY-breaking parameters.
Regarding the latter aspect, let us compare $\delta_{B_\kappa}
\bmm_\nu$ and $\delta_{C_\kappa} \bmm_\nu$ with the tree-level terms
$\bmm^{(\kappa)}_\nu$ and $\bmm^{(\tilde{B}_\kappa)}_\nu$ of
eqs.~(\ref{mnuk}) and (\ref{mnubtk}). Suppose there are two
SUSY-breaking mass scales $\tilde{m}$ and $\tilde{m}_\kappa$, such
that sleptons and gauginos have masses of order $\tilde{m}$, while
the SUSY-breaking terms in eq.~(\ref{K2}) scale as\footnote{We
replace bold characters with unbolded ones whenever we discuss order
of magnitude estimates for some quantity.} $B_\kappa \sim
\tilde{B}_\kappa \sim \kappa \tilde{m}_\kappa$, $C_\kappa \sim
\kappa \tilde{m}_\kappa^2$. Then, the relative corrections
$\delta_{B_\kappa} m_\nu/ m^{(\kappa)}_\nu$ and $\delta_{C_\kappa}
m_\nu/ m^{(\tilde{B}_\kappa)}_\nu$ are of order $10^{-3} \,
\tilde{m}_\kappa/\tilde{m}$. In particular, they are negligible for
$\tilde{m}_\kappa \sim \tilde{m}$ while they can be ${\cal O}(10
\%)$ for $\tilde{m}_\kappa \sim 10^2 \, \tilde{m}$, as in the case
of the explicit model presented in Section~4.


\section{Type II seesaw realizations}

At this point, a natural question arises about the possible origin
of the $d=6$, $\Delta L=2$ effective operators discussed, so far, in
a general way. By considering simple scenarios like the type
I/II/III seesaw mechanisms (which generate the familiar $d=5$
superpotential operator at the tree level), we immediately realize
that the type II framework is the natural one in which those $d=6$
operators emerge. As a matter of fact, the tree-level exchange of
type I or type III mediators leads to $\Delta L=0$ K\"ahler
operators of the form $|H_2 L|^2$, whereas the type II mediators
induce both $\Delta L=0$ and $\Delta L=2$ operators.


\subsection{Type II in the SUSY limit}

The type II seesaw mechanism is realized through
the exchange of $SU(2)_W$ triplet states $T =(T^0, T^+, T^{++})$
and $\bar{T}=(\bar{T}^0, \bar{T}^-, \bar{T}^{--})$
in a vector-like $SU(2)_W\times U(1)_Y$ representation, $T\sim
(3,1)$, $\bar{T} \sim (3,-1)$. The relevant superpotential terms are:
\be
\label{L-T}
W \supset
\frac{1}{\sqrt{2}}\bY^{ij}_{T} L_i T L_j + \frac{1}{\sqrt{2}}\la_1
H_1 T H_1 +
\frac{1}{\sqrt{2}} \la_2 H_2 \bar{T} H_2 +
M_T T \bar{T}\,,
\ee
where $\bY^{ij}_{T}$ is a  $3 \times 3$ symmetric matrix,
$\la_{1,2}$ are dimensionless couplings and $M_T$ is the (SUSY)
triplet mass. Neutrino masses are usually generated through the
$d=5$ effective superpotential term $\frac{\la_2}{2M_T} \bY^{ij}_{T}
(L_i H_2) (L_j H_2)$, which is the leading $\Delta L=2$ operator
emerging from the exchange of the triplet states. We assume this
contribution to be strongly suppressed (absent) by a very small
(vanishing) value of $\la_2$. This can be either imposed {\em ad
hoc} or justified by symmetry arguments, like those presented in
Section~2.1. For instance, the smallness of $\la_2$ could be related
to the smallness of $\mu$ (\eg, if we assign zero PQ charge to
$\la_1$ and $M_T$, then we expect $\la_2 \sim \mu^2/\Lambda_*^2$
since the PQ charge of $\la_2$ is twice that of $\mu$). In a NMSSM
framework, simple $Z_3$ assignments can forbid $\la_2$ and allow the
remaining terms in eq.~(\ref{L-T}) (\eg, one can assign $Z_3$ charge
$-1/3$ to $\bar{T}$ and $1/3$ to all the other fields).

Once $\la_2$ is disregarded, the leading $\Delta L=2$ operator is
precisely the K\"ahler operator $(H^\dagger_1 L)^2$. Indeed, we can
integrate out the heavy states by imposing $\partial W/\partial T=0$
and plugging the expression of $\bar{T}$ into the canonical K\"ahler
term $\int \!  d^4 \theta \, \bar{T}^\dagger \bar{T}$. As a result,
the following $\Delta L=2$ effective operator is generated at the
scale $M_T$:
\be
K_{\rm eff} \supset \frac{\la^\ast_1}{2 |M_T|^2} \bY^{ij}_T
(H^\dagger_1 L_i )(H^\dagger_1 L_j ) + h.c. \, ,
\label{d6k}%
\ee
along with other $\Delta L=0$ operators. The above term can be
matched to the SUSY part of eq.~(\ref{K1}) [or eq.~(\ref{K2})]
through the identification
\be
\label{ident} \bkappa = \la_1^* \bY_T  \;\;\; , \;\;\;
M^2=|M_T|^2\,.
\ee
The resulting contribution to neutrino masses is the tree-level term
$\bmm^{(\kappa)}_\nu$ of eq.~(\ref{mnuk}). Its diagrammatic
interpretation is shown in the left diagram of Fig.~\ref{f4}, which is the
explicit realization of the left diagram of Fig.~\ref{f2}. The
appearance of such a contribution to neutrino
masses through triplet exchange was noticed in \cite{AR} (and also
in \cite{KLL}, where $\mu$ was generated by the VEV of a SM singlet
$S$, charged under an extra $U(1)$).


\subsection{Type II with broken SUSY}

We now address the question on how the SUSY-breaking operators of
eqs.~(\ref{K1}) or (\ref{K2}) can arise in the type II seesaw
framework. To this purpose, we distinguish three SUSY-breaking
scenarios, depending on the ordering of the SUSY-breaking mediation
scale $M_S$ and the triplet mass $M_T$: {\it i}) $M_S < M_T$; {\it
ii}) $M_S > M_T$; {\it iii}) $M_S = M_T$.

\vskip .5 cm

{\it i}) $M_S < M_T$. Suppose that SUSY breaking is mediated at a
low scale $M_S < M_T$ by a messenger sector coupled to the MSSM
states through gauge interactions only (pure gauge mediation
\cite{gmed,gmedrev}). In this case, SUSY-breaking gaugino and
(flavour blind) sfermion masses of order $\tilde{m}$ arise at $M_S$
through loop diagrams, while trilinear couplings are mainly
generated below $M_S$ by gaugino mass terms in the RGEs. Moreover,
the flavour structure of the sfermion masses and trilinear couplings
emerges through RGEs only and is entirely controlled by the Yukawa
matrices. Regarding our $\Delta L=2$ operators of eq.~(\ref{K2}), in
this scenario only the SUSY one with coefficient $\bkappa$ exists
above $M_S$ (it is generated at $M_T$). The SUSY-breaking parameters
${\bB}_\kappa$, ${\tilde \bB}_\kappa $ and $\bC_\kappa$ receive
two-loop finite contributions at $M_S$, proportional to $\bkappa$,
and important corrections are generated below $M_S$ by gaugino or
scalar masses through RGEs. At low-energy, the flavour structure of
such parameters is controlled by $\bkappa$ and $\bY_e$, while their
expected size is ${\bB}_\kappa ,{\tilde \bB}_\kappa \ll \bkappa
\tilde{m}$ and $\bC_\kappa \sim \bkappa \tilde{m}^2$. Therefore, in
such a scenario the dominant source of neutrino masses is,
generically, the SUSY contribution $\bmm^{(\kappa)}_\nu$ of
eq.~(\ref{mnuk}), while the SUSY-breaking contributions
$\bmm^{(\tilde{B}_\kappa)}_\nu$, $\delta_{B_\kappa} \bmm_\nu$ and
$\delta_{C_\kappa} \bmm_\nu$ of eqs.~(\ref{mnubtk}),
(\ref{deltambk}) and (\ref{deltamck}) are subleading. We also remark
that these comments do not rely on the specific realization of
$\bkappa$ described in Section~3.1, but hold in general for
gauge-mediated SUSY breaking at scales $M_S < M$.


\begin{figure}[]

\begin{center}
\begin{picture}(120,80)(-60,-40)
\ArrowLine(-50,-25)(0,-25)
\ArrowLine(50,-25)(0,-25)
\DashArrowLine(0,40)(0,-25){3}
\DashArrowLine(0,40)(-40,60){3}
\DashArrowLine(0,40)(20,50){3}
\DashArrowLine(40,60)(20,50){3}
\Text(-50,-35)[]{\small $L$}
\Text(50,-35)[]{\small $L$}
\Text(2,-34)[]{\small $\bY_T$}
\Text(-10,12)[]{\small $T$}
\Text(-45,50)[]{\small $H_1$}
\Text(15,35)[]{\small $F_{H_1}$}
\Text(45,50)[]{\small $H_2$}
\Text(20,50)[]{\small $+$}
\Text(15,55)[]{\small $\mu$}
\Text(0,50)[]{\small $\lambda_1^*$}
\Text(25,10)[]{\small $1/|M_T|^2 $}
\end{picture}
\hglue 1.0 cm
\begin{picture}(120,80)(-60,-40)
\ArrowLine(-50,-25)(0,-25)
\ArrowLine(50,-25)(0,-25)
\DashArrowLine(0,5)(0,-25){3}
\DashArrowLine(0,5)(0,25){3}
\DashArrowLine(0,40)(0,25){3}
\DashArrowLine(0,40)(-40,60){3}
\DashArrowLine(0,40)(40,60){3}
\BCirc(0,5){3}
\Text(0,5)[]{\small $\times$}
\Text(-50,-35)[]{\small $L$}
\Text(50,-35)[]{\small $L$}
\Text(2,-34)[]{\small $\bY_T$}
\Text(-10,-10)[]{\small $T$}
\Text(-10,18)[]{\small $\bar{T}$}
\Text(-10,33)[]{\small $F_T$}
\Text(-45,50)[]{\small $H_1$}
\Text(45,50)[]{\small $H_1$}
\Text(23,5)[]{\small $M_T^* B_T^*$}
\Text(25,-13)[]{\small $1/|M_T|^2 $}
\Text(20,25)[]{\small $1/M_T^* $}
\Text(0,50)[]{\small $\lambda_1^*$}
\end{picture}
\hglue 1.0 cm
\begin{picture}(120,80)(-60,-40)
\ArrowLine(-50,-25)(0,-25)
\ArrowLine(50,-25)(0,-25)
\DashArrowLine(0,40)(0,-25){3}
\DashArrowLine(0,40)(-40,60){3}
\DashArrowLine(0,40)(40,60){3}
\BCirc(0,40){3}
\Text(0,40)[]{\small $\times$}
\Text(-50,-35)[]{\small $L$}
\Text(50,-35)[]{\small $L$}
\Text(2,-34)[]{\small $\bY_T$}
\Text(-10,12)[]{\small $T$}
\Text(-45,50)[]{\small $H_1$}
\Text(45,50)[]{\small $H_1$}
\Text(0,50)[]{\small $A_1^*$}
\Text(25,10)[]{\small $1/|M_T|^2 $}
\end{picture}
\end{center}
\vspace{-0.5 cm}
\caption{\small Tree-level contributions to neutrino masses from
heavy triplet exchange.}
\label{f4}
\end{figure}
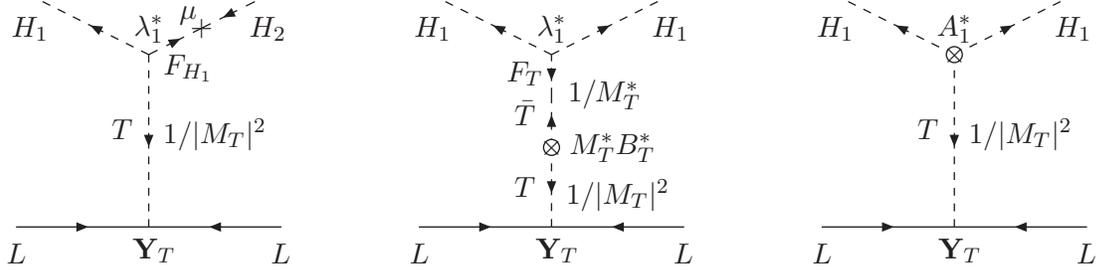

\vskip .5 cm

{\it ii}) $M_S > M_T$. Suppose that SUSY-breaking terms are
generated at a high scale $M_S>M_T$ through, \eg, gravity or gauge
mediation. This implies that at $M_T$ all the MSSM and triplet
fields generically have SUSY-breaking mass parameters. In
particular, we can write those related to the triplets by replacing
$M_T \rightarrow M_T(1 -\theta^2 B_T)$, $\la_1 \rightarrow \la_1
-\theta^2 A_1$, $\bY_T \rightarrow \bY_T -\theta^2  \bA_T$ in the
superpotential terms (\ref{L-T}) and $\int \!  d^4 \theta \, [
T^\dagger T + \bar{T}^\dagger \bar{T}] \rightarrow \int \!  d^4
\theta \,[(1 - \theta^2 \bar{\theta}^2 m^2_T) T^\dagger T + (1 -
\theta^2 \bar{\theta}^2 m^2_{\bar{T}})\bar{T}^\dagger \bar{T} ]$ in
the canonical K\"ahler part. As previously done, we can integrate
out the triplet states, including now such SUSY-breaking effects. In
this way, we obtain all the $\Delta L=2$ effective operators of
eq.~(\ref{K2}), namely, both the SUSY one with $\bkappa = \la_1^*
\bY_T$ [eqs.~(\ref{d6k}) and (\ref{ident})] and the SUSY-breaking
ones with
\be
\label{bks}
\bB_\kappa = \la_1^* (\bY_T B_T - \bA_T) \;\;\; ,
\;\;\;\;\;
{\tilde \bB}_\kappa = (\la_1^* B_T^* -A_1^*) \bY_T  \;\;\; ,
\ee
\be
\label{ck} \bC_\kappa = (\la_1^* B_T^* -A_1^*)  (\bY_T B_T - \bA_T)
- \la_1^* \bY_T m^2_{\bar{T}}\,.
\ee
We recall that ${\tilde \bB}_\kappa$ contributes to neutrino masses
at the tree-level [see $\bmm^{(\tilde{B}_\kappa)}_\nu$ in
eq.~(\ref{mnubtk})]. Its diagrammatic origin from triplet exchange
is shown in the middle and right diagrams of Fig.~\ref{f4}, which
are the explicit realization of the right diagram of Fig.~\ref{f2}.
The diagrammatic interpretation of $\bB_\kappa$ and $\bC_\kappa$ is
straightforward. The relative size of the parameters which
contribute to neutrino masses is model dependent, while their
flavour structure exhibit remarkable features, as a consequence of
the type II seesaw mechanism. In particular, $\bkappa$ and ${\tilde
\bB}_\kappa$ (which generate the leading contributions to neutrino
masses) are aligned in flavour space at $M_T$, as both are
proportional to $\bY_T$. Some misalignment is induced by $\bY_e$ and
$\bA_e$ through RGEs [see eqs.~(\ref{rgek}) and (\ref{rgebkt})]. On
the other hand, $\bB_\kappa$ and $\bC_\kappa$ (which contribute to
$\bmm_\nu$ through low-energy threshold corrections) owe their
flavour dependence to $\bY_T$ and $\bA_T$ already at $M_T$, and
acquire further structure via RGEs. Nevertheless, if the mechanism
of SUSY-breaking mediation at $M_S$ is flavour blind, like in
minimal gravity- or gauge-mediated models, all flavour structures
are controlled by $\bY_T$ and $\bY_e$.

\vskip .5 cm

{\it iii}) $M_S = M_T$. This case deserves special attention, and we
will discuss it thoroughly in the next section.


\section{Seesaw mediators as SUSY-breaking messengers}

We have seen that the type II scenario provides a natural framework
to induce neutrino masses through K\"ahler operators. So far, the
mechanisms that mediate SUSY breaking and lepton number violation
have been kept distinct. Now we discuss an appealing scenario in
which such mechanisms are unified, namely, the seesaw mediators are
identified with the SUSY-breaking ones. This idea was proposed and
thoroughly explored by two of us in \cite{JR}, where neutrino masses
were generated through an effective $d=5$ superpotential in a type
II scenario. Further developments were presented in \cite{seesawmed}
with either type II or type III mediators. We now aim at extending
such an approach to our framework with $d=6$ K\"ahler operators. We
present a minimal scenario in which a single SUSY-breaking source
determines the sparticle spectrum and plays a major r\^ole in
generating neutrino masses. Moreover, the magnitude and flavour
structure of all mass parameters are closely correlated.

\subsection{SUSY-breaking mediation}

We start by identifying the type II triplets with SUSY-breaking
mediators and embed them in a minimal messenger sector which, in
order to generate the gluino mass, should also include coloured
fields. In addition, we require that perturbative unification of
gauge couplings be preserved and that all messenger masses be of the
same order. This implies that the messenger sector should have the
same total Dynkin index $N$ for each subgroup of $SU(3)_C \times
SU(2)_W \times U(1)_Y$. One way to realize this is to embed the
$SU(2)_W$ triplets $T$ and $\bar{T}$ into complete $SU(5)$
representations\footnote{ $SU(5)$ extensions of the type II seesaw
with gravity-mediated SUSY breaking have been discussed in
\cite{AR,porod,borzyam,FJ}.}, like in standard gauge mediation. The
simplest such embedding is $T \subset 15$ and $\bar{T} \subset
\ov{15}$, which has $N=7$ \cite{JR}. Alternatively, the messenger
sector could have the same $N$ for each group factor even without
filling unified multiplets~\cite{mart,magic}. Since $T$ and
$\bar{T}$ have $SU(2)$ index $N_2=4$, we are constrained to $N \geq
4$. In particular, we can look for a minimal messenger sector with
$N=4$. One possible choice relies on adding a pair of $SU(3)_C$
triplets $(3,1,-1/3)+(\bar{3},1,+1/3)$ and an $SU(3)_C$ adjoint
$(8,1,0)$ to the $T+\bar{T}$ pair\footnote{This is one of the
`magic' combinations listed in \cite{magic}.}. The octet can also be
replaced by three pairs of coloured triplets, \ie \, $(8,1,0)
\rightarrow 3 \times [(3,1,0)+(\bar{3},1,0)]$. These are the only
two possibilities with $N=4$ and no exotic charges.

Following the standard parametrization of minimal gauge mediation
\cite{gmed,gmedrev}, we write messenger mass terms as $W \supset
\xi_i X \Phi_i \bar{\Phi}_i$ (or $\xi_i X \Phi_i^2/2$ for real
representations), where $\langle X|_0 \rangle = v_X$ and $\langle
X|_{\theta^2} \rangle = F_X$. Thus, the messengers have SUSY masses
$M_i = \xi_i v_X$ and a common $B$-parameter $\Lambda = F_X/v_X$,
which is usually named effective SUSY-breaking scale. Hereafter, we
set $\Lambda \equiv - B_T$ (consistently with our notation) 
and assume that all $\xi_i$ are of the same
order, so that we can deal with a common messenger scale $M_i \sim
M_T$.

The MSSM SUSY-breaking parameters are generated at the quantum level
by a messenger sector of the type described above, coupled to the
MSSM fields through both gauge and Yukawa interactions\footnote{In
our case, the relevant Yukawa couplings are $\la_1$ and $\bY_T$. We
neglect the effect of other Yukawa couplings which may involve
messenger fields (see \eg \, \cite{JR}).}. At the one-loop level,
the gaugino masses $M_a$, the Higgs $B$-term $B_H$ and the trilinear
terms $\bA_x$ are:
\be
\label{mabh}
M_a =  -\frac{N B_T }{16 \pi^2}\, g_a^2
\;\;\;\; , \;\;\;\;
B_H  = \frac{3 B_T }{16 \pi^2}\, |\la_1|^2  \; ,
\ee
\be
{\bA}_e =  \frac{3 B_T}{16 \pi^2} \bY_e (\bY^\dagger_T \bY_T
+|\la_1|^2 ) \;\; , \;\; {\bA}_d = \frac{ 3 B_T}{16 \pi^2} \bY_d
|\la_1|^2 \;\; , \;\; {\bA}_u = 0 \; ,
\ee
where $g_1^2=(5/3)g'^2$ and $g_2^2=g^2$. Non-vanishing ${\cal
O}(B_T^2)$ contributions for the squared scalar masses arise at the
two-loop level:
\bea
\label{mlsoft}
\!\!\!\!\bmm^2_{\tl}& =& \left(\frac{|B_T|}{16 \pi^2}\right)^2 \left[
N \left( \frac{3}{10} g^4_1 + \frac{3 }{2} g^4_2 \right)
- \left(\frac{27}{5} g^2_1
+21 g^2_2\right)\!\bY^\dagger_T\bY_T
+ 3 |\la_1|^2 (\bY^\dagger_T \bY_T-
\bY^\dagger_e \bY_e)
\right. \no \\
&& \phantom{xxxxxxx}
\left.
+ 3\, \bY^\dagger_T (\bY^\dagger_e \bY_e)^T \bY_T
+ 18 \, (\bY^\dagger_T\bY_T)^2
+ 3 \, \bY^\dagger_T\bY_T {\rm Tr}(\bY^\dagger_T\bY_T)
\right]\,, \\
\!\! \bmm^2_{\te}& =& \left(\frac{|B_T|}{16 \pi^2}\right)^2
\left[N \left( \frac{6 }{5} g^4_1 \right)
- 6 \, \bY_e( \bY^\dagger_T\bY_T+
 |\lambda_1|^2 )\bY^\dagger_e \right]\,,
\\
\!\!\bmm^2_{\tq}& = &\left(\frac{|B_T|}{16 \pi^2}\right)^2
\left[ N \left( \frac{1}{30} g^4_1 + \frac{3 }{2} g^4_2
+ \frac{8 }{3} g^4_3 \right)
-3  |\la_1|^2 \, \bY^\dagger_d \bY_d
\right]\,, \\
\!\!\bmm^2_{\tu} &= &\left(\frac{|B_T|}{16 \pi^2}\right)^2
\left[N \left( \frac{8 }{15}
g^4_1 +\frac{8 }{3} g^4_3 \right) \right]\,,\no \\
\!\!\bmm^2_{\td}& = &\left(\frac{|B_T|}{16 \pi^2}\right)^2
\left[N \left(\frac{2 }{15} g^4_1 +\frac{8}{3} g^4_3 \right)
- 6 |\la_1|^2 \, \bY_d \bY^\dagger_d  \right]\,, \\
\!\! m^2_{H_2}& = &\left(\frac{|B_T|}{16 \pi^2}\right)^2\left[
 N \left( \frac{3 }{10} g^4_1 +   \frac{3 }{2}  g^4_2 \right)
\right]\,,  \\
\label{mh1}
\!\! m^2_{H_1} &= &\left(\frac{|B_T|}{16 \pi^2}\right)^2\left[
 N \left( \frac{3 }{10} g^4_1 +   \frac{3 }{2}  g^4_2 \right)
- \left(\frac{27}{5} g^2_1
+21 g^2_2\right)|\la_1|^2
+  21 |\la_1|^4
\right. \no \\
&& \phantom{xxxxxxx} \left. + 3  |\la_1|^2 \, {\rm Tr}(
\bY^\dagger_T\bY_T + \bY^\dagger_e\bY_e + 3\bY^\dagger_d\bY_d ) -3\,
{\rm Tr}(\bY^\dagger_T\bY_T \bY^\dagger_e\bY_e) \right]\, .
\eea
The above results follow from simple changes 
(including a correction of the $M_a$ sign) in the formulae of
\cite{JR}, which were derived by applying the method\footnote{This
method provides the leading terms of a power expansion in
$|B_T/M_T|^2$, which we assume to be $\ll 1$. The latter condition
also allows us to neglect: {\it i}) corrections such as $\delta
\bmm^2_{\tl} \simeq -\bY^\dagger_T\bY_T |B_T|^4/(32 \pi^2 |M_T|^2)$
and $\delta m^2_{H_1} \simeq -|\la_1|^2 |B_T|^4/(32 \pi^2 |M_T|^2)$,
which are the leading one-loop contributions to scalar masses; {\it
ii}) quartic terms like $-|\la_1 B_T/M_T|^2 |H_1|^4/2$, induced in
the scalar potential by the tree-level exchange of the triplets;
{\it iii}) four-lepton operators generated by the tree-level
exchange of the triplets, which can contribute to LFV processes such
as $\mu \to 3 e$ (see, \eg, \cite{triplfv,abada,lfvcern}).} of
\cite{gr}. Eqs.~(\ref{mabh})-(\ref{mh1}) form the complete set of
boundary conditions at $M_T$ for the SUSY-breaking parameters, which
must be subsequently renormalized down to low energies. Notice that
the flavour structures of $\bA_e$, $\bmm^2_\tl$ and $\bmm^2_{\te}$
are controlled by $\bY_T$ and $\bY_e$, which in turn are closely
related to the low-energy lepton masses and mixing angles. Such
minimal LFV properties are a characteristic feature of the SUSY type
II seesaw \cite{AR,JR}. Clearly, our scenario possesses the property
of minimal flavour violation~\cite{mfv} in both the quark and lepton
sectors. The former is controlled by the usual spurions $\bY_u$ and
$\bY_d$, while in the latter the matrices $\bY_T$ and $\bY_e$ are
the spurions of the (minimal) lepton flavour symmetry $SU(3)_L
\times SU(3)_{E^c}$, under which $\bY_T \sim (\bar{6},1)$ and $\bY_e
\sim (\bar{3},\bar{3})$. All leptonic quantities depend on invariant
combinations of such spurions. For instance, the aforementioned
symmetry allows $\bmm^2_{\tl}$ to contain structures like
$\bY^\dagger_T \bY_T$ and $\bY^\dagger_e \bY_e$ at the quadratic
level and $(\bY^\dagger_T \bY_T)^2$, $\bY^\dagger_T (\bY^\dagger_e
\bY_e)^T \bY_T$, $(\bY^\dagger_T \bY_T)(\bY^\dagger_e \bY_e)$,
$(\bY^\dagger_e \bY_e)(\bY^\dagger_T \bY_T)$, $(\bY^\dagger_e
\bY_e)^2$ at the quartic one. In fact, all such combinations are
present in $\bmm^2_{\tl}$ at low energy, since some appear at $M_T$
[eq.~(\ref{mlsoft})] and others are induced through RGEs. Finally,
we remark that there is a phase alignment among $M_a$, $B_H$ and
$\bA_x$, induced by the common factor $B_T$. As a consequence, the
one-loop sfermion/gaugino/higgsino contributions to the
electric-dipole moments are strongly suppressed.

\subsection{Neutrino masses}

Upon decoupling the triplets, the MSSM SUSY-breaking masses are
generated through finite radiative effects
[eqs.~(\ref{mabh})--(\ref{mh1})], while the $\Delta L=2$
SUSY-breaking parameters $\bB_\kappa$, ${\tilde \bB}_\kappa$ and
$\bC_\kappa$ arise at the tree level. The latter have a very simple
form, namely that of eqs.~(\ref{bks}) and (\ref{ck}) with vanishing
$A_1$, $\bA_T$ and $m^2_{\bar{T}}$:
\be
\label{simple}
\bB_\kappa = B_T \, \bkappa
\;\;\; ,
\;\;\;\;\;
{\tilde \bB}_\kappa = B_T^* \, \bkappa
\;\;\; , \;\;\;\;\;
\bC_\kappa =|B_T|^2 \, \bkappa\,,
\ee
where $ \bkappa =  \la_1^* \bY_T$ [eq.~(\ref{ident})]. These
alignment relations hold at the scale $M_T$, and remain also valid
to a very good approximation after RG evolution, which is dominated
by the homogeneous terms. Indeed, the non-homogeneous terms in the
RGEs (\ref{rgebk})-(\ref{rgeck}) are proportional to MSSM
SUSY-breaking parameters, which are loop-suppressed with respect to
$B_T$. So, we can apply eq.~(\ref{simple}) also at low scales, as
long as we take into account the RG running of $\bkappa$
[eq.~(\ref{rgek})]. The relative size of $\bB_\kappa$, ${\tilde
\bB}_\kappa$ and $\bC_\kappa$ is also completely determined and,
therefore, it is simple to evaluate and compare the corresponding
contributions to neutrino masses.

Inserting now the expression of $\tilde{\bB}_\kappa$ from
eq.~(\ref{simple}) into the general eqs.~(\ref{mnu})-(\ref{mnubtk}),
we can write the tree-level contribution to the neutrino mass matrix
as:
\be
\label{mnusum} \bmm_\nu = \bkappa \, ( B_T^* +  2\, \mu \tan\beta)
\cos^2 \! \beta \; \frac{ v^2}{|M_T|^2} \, .
\ee
Although we have not specified the mechanism which generates $\mu$,
we note that $\mu$ is suppressed with respect to $B_T$ since the
conditions of electroweak symmetry breaking (EWSB) connect $\mu$ with other
SUSY-breaking parameters, whose size ${\tilde m}$ is related to
$B_T$ by a loop factor. Hence, we expect $\tilde{\bB}_\kappa = B_T^*
\, \bkappa$ to be the {\em dominant source of neutrino masses} in
the present scenario. The contribution proportional to $\mu$ may
become comparable to such leading term only for large values of
$\tan\beta$.

Regarding the finite quantum corrections to $\bmm_\nu$ discussed in
Section 2.5, we conclude that $\delta_\kappa \bmm_\nu$ is very small
[see eq.~(\ref{deltamk}) and related comments]. The other
contribution $\delta \bmm_\nu = \delta_{B_\kappa} \bmm_\nu +
\delta_{C_\kappa} \bmm_\nu$ induced by $\bB_\kappa$ and $\bC_\kappa$
is more interesting. From eqs.~(\ref{deltambk}) and (\ref{deltamck})
we get:
\be
\delta \bmm_\nu \simeq
 \frac{1}{2 N} \left( f_{L2} + \frac{3}{5} f_{L1} \right) \bmm_\nu
- \frac{1}{2 N} \left( h_{L2} + \frac{3}{5} h_{L1} \right)
(\bmm_\nu \, \bD_L + \bD_L^T \bmm_\nu )\,,
\ee
which is directly related to the tree-level term $\bmm_\nu$ of
eq.~(\ref{mnusum}). We also note that the parametric
loop-suppression of $\delta \bmm_\nu$ has disappeared, since the
low-energy loop factor in eqs.~(\ref{deltambk}) and (\ref{deltamck})
has been compensated by the inverse loop factor in $B_T/M_a$ [see
eq.~(\ref{mabh})]. There is still a residual numerical suppression,
which depends on the messenger index $N$ and on the low-energy
values of ${\tilde m}_L^2/|M_a|^2$ entering $f_{La}$ and $h_{La}$.
For instance, the leading correction (which is always aligned with
$\bmm_\nu$) is $\delta \bmm_\nu \simeq 0.1 \, \bmm_\nu $ ($ 0.07
\, \bmm_\nu$) for $N = 4$ ($N=7$). Furthermore, even the small
flavour-dependent term induced by $\bD_L$ does not exhibit an
independent structure, since both $\bD_L$ and $\bmm_\nu$ are
controlled by the basic lepton flavour spurions $\bY_T$ and $\bY_e$.
This confirms the minimal LVF properties of the present scenario. We
also recall that, in general, $\bD_L$ also induces a misalignment
between $\bY_e$ and the charged lepton mass matrix ${\cal M}_e$
through $\tan\beta$-enhanced threshold corrections \cite{bkbr}. In
our case, the resulting effects on the lepton mixing matrix $\bU$
are again controlled by our minimal LFV structure and are
numerically small.

Finally, it is worth mentioning that the $\Delta L=2$ {\em
sneutrino} mass matrix of eq.~(\ref{sneu}) is directly linked to
$\bmm_\nu$ through $\delta \bmm^2_{\tilde{\nu}} = B_T \bmm_\nu$,
which is the same relation found in the $d=5$ realization of
\cite{JR}. However, in both cases the conditions for the
observability of sneutrino-antisneutrino oscillations \cite{grhab}
are not fulfilled (despite the large enhancement factor $B_T/{\tilde
m}_L $) since the oscillation frequency $\Delta m_{\tilde{\nu}} \sim
\delta \bmm^2_{\tilde{\nu}}/{\tilde m}_L \sim (B_T/{\tilde m}_L) \,
m_\nu \sim 10^2 \, m_\nu $ is much smaller than the sneutrino decay
width. Indeed, since two body decay channels like $\tilde{\nu}
\rightarrow \nu \, \tilde{\chi}^0_1$ are open, we have
$\Gamma_{\tilde{\nu}} \sim 10^{-3} \, {\tilde m}_L $, implying
$\Delta m_{\tilde{\nu}}/\Gamma_{\tilde{\nu}} \lsim 10^{-7}$.


\subsection{Phenomenological viability, MSSM spectrum and LHC searches}

The above scenario has a small number of free parameters, namely
$M_T$, $B_T$, $\lambda_1$ and the messenger index $N$. Once these
are fixed\footnote{Hereafter the parameters $M_T$, $B_T$ and
$\lambda_1$ are taken as real, without loss of generality.}, the
remaining parameters $\bY_T$, $\tan\beta$ and $\mu$ are determined
by the low-energy neutrino data and by requiring proper EWSB. 
Concerning neutrino data, we recall that the
neutrino mass matrix $\bmm_\nu$ is related to the low-energy
observables as $\bmm_\nu = \bU^* \bmm^{D }_\nu \bU^\dagger$, where  
$\bmm^{D }_\nu = {\rm diag}(m_1,m_2,m_3)$, 
$m_a$ are the neutrino masses and $\bU$ is the
lepton mixing matrix\footnote{ We use the standard parametrization
$\bU=\bV(\theta_{12},\theta_{23},\theta_{13}, \delta) \cdot {\rm
diag}(1,e^{i \phi_1},e^{i \phi_2})$. In our numerical analysis we
will use the best-fit values for the neutrino parameters $\Delta
m^2_{21}=(m^2_2-m^2_1)= 7.65 \times 10^{-5} {\rm eV}^2$, $|\Delta
m^2_{31}|=|m^2_3-m^2_1|= 2.4 \times 10^{-3} {\rm eV}^2$, $\sin^2 \!
\theta_{12}=0.3$, $\sin^2 \! \theta_{23}=0.5$ and the upper bound
$s_{13} =\sin \theta_{13} <0.2$ \cite{nuexp,nupheno}.}. Several
other observables are predicted, such as sparticle and Higgs masses,
as well as LFV decay rates.

Before presenting a numerical analysis, we can already infer some
information about the allowed parameter space by considering the
parametric dependence of the neutrino mass in eq.~(\ref{mnusum}),
$m_\nu \sim Y_T \la_1 \! \cos^2 \! \beta B_T v^2/M_T^2$. For $B_T
\lsim 10^5$ GeV (which leads to a superpartner spectrum below a few
TeV, within the reach of the LHC) and $Y_T$, $\la_1 \lsim 1$, a
neutrino mass scale $m_\nu \sim 0.1~{\rm eV}$ requires $M_T \lsim
10^9~\gev$. This {\em upper} bound on $M_T$ implies a non-trivial
constraint on the messenger sector. Indeed, suppose we choose the
simplest grand unified embedding with $T \subset 15$ and $\bar{T}
\subset \ov{15}$ ($N=7$). In this case, the {\em lower} bound on
$M_T$ compatible with one-loop gauge coupling unification is $M_T
\gsim 10^7$ GeV. At the two-loop level, we find the stronger
constraint $M_T \gsim 5 \times 10^8$ GeV. We have explored the
parameter space for this scenario, taking into account the bounds on
the lightest Higgs mass $m_h$ \cite{hlep} and on rare LFV decays
\cite{MEGA,babar,belle}. The outcome is a certain tension between
these constraints and that on perturbative gauge coupling
unification. In other words, a messenger sector with heavy states in
$15+\ov{15}$ ,which is perfectly compatible with the $d=5$
realization of neutrino masses \cite{JR}, is only marginally
compatible with the $d=6$ scenario proposed here. Hence, we will
present quantitative results for a smaller messenger sector, namely,
a minimal one with $N=4$. As mentioned in Section~4.1, this can be
realized, \eg, by adding $SU(3)_C$ triplets
$(3,1,-1/3)+(\bar{3},1,+1/3)$ and an adjoint $(8,1,0)$ to the
$T+\bar{T}$ pair. Perturbative gauge coupling unification is no
longer a problem in this case, since it can be achieved with
messenger masses as low as $10^5$ GeV.

\begin{figure}[p]
\begin{center}
\begin{tabular}{cc}
\hspace*{-0.1cm}\includegraphics[width=16.5cm,height= 8.5cm]{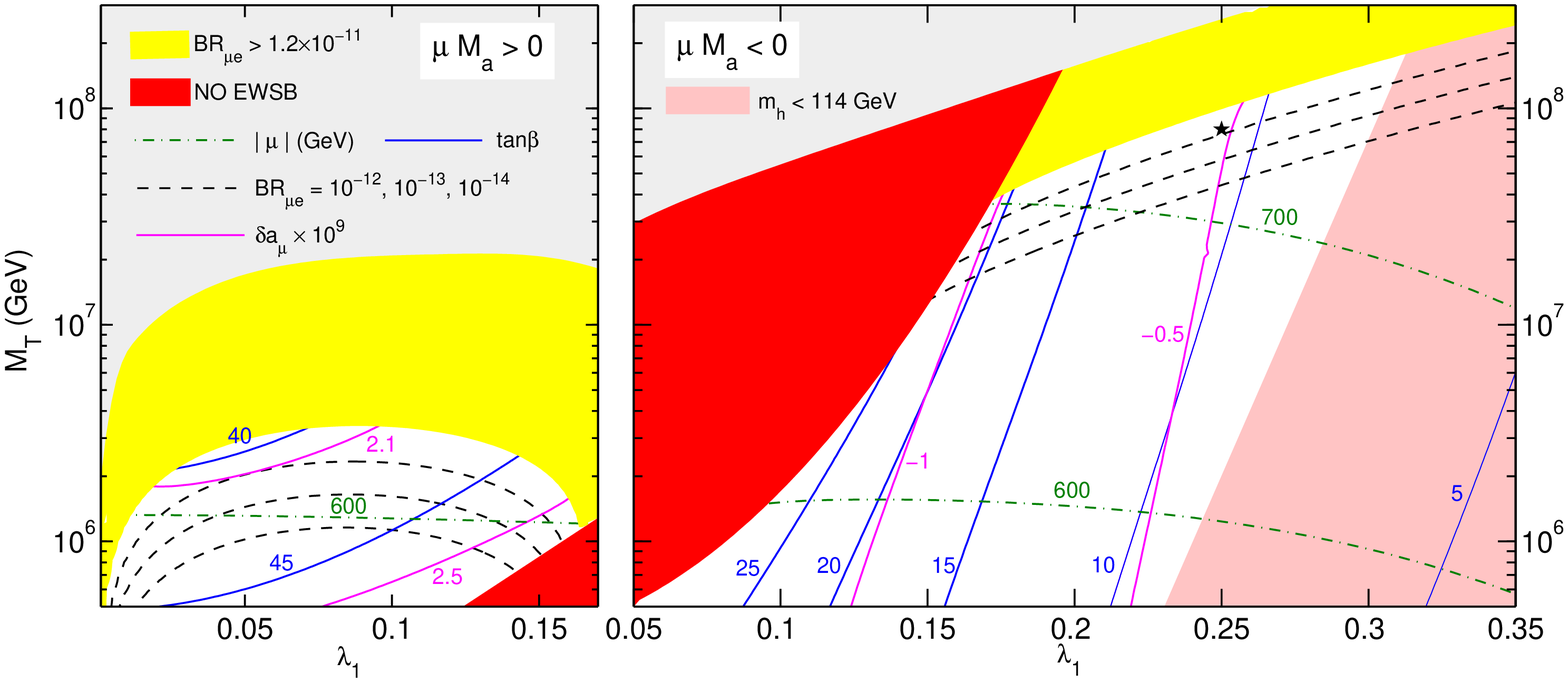} \\
\includegraphics[width=9.5cm]{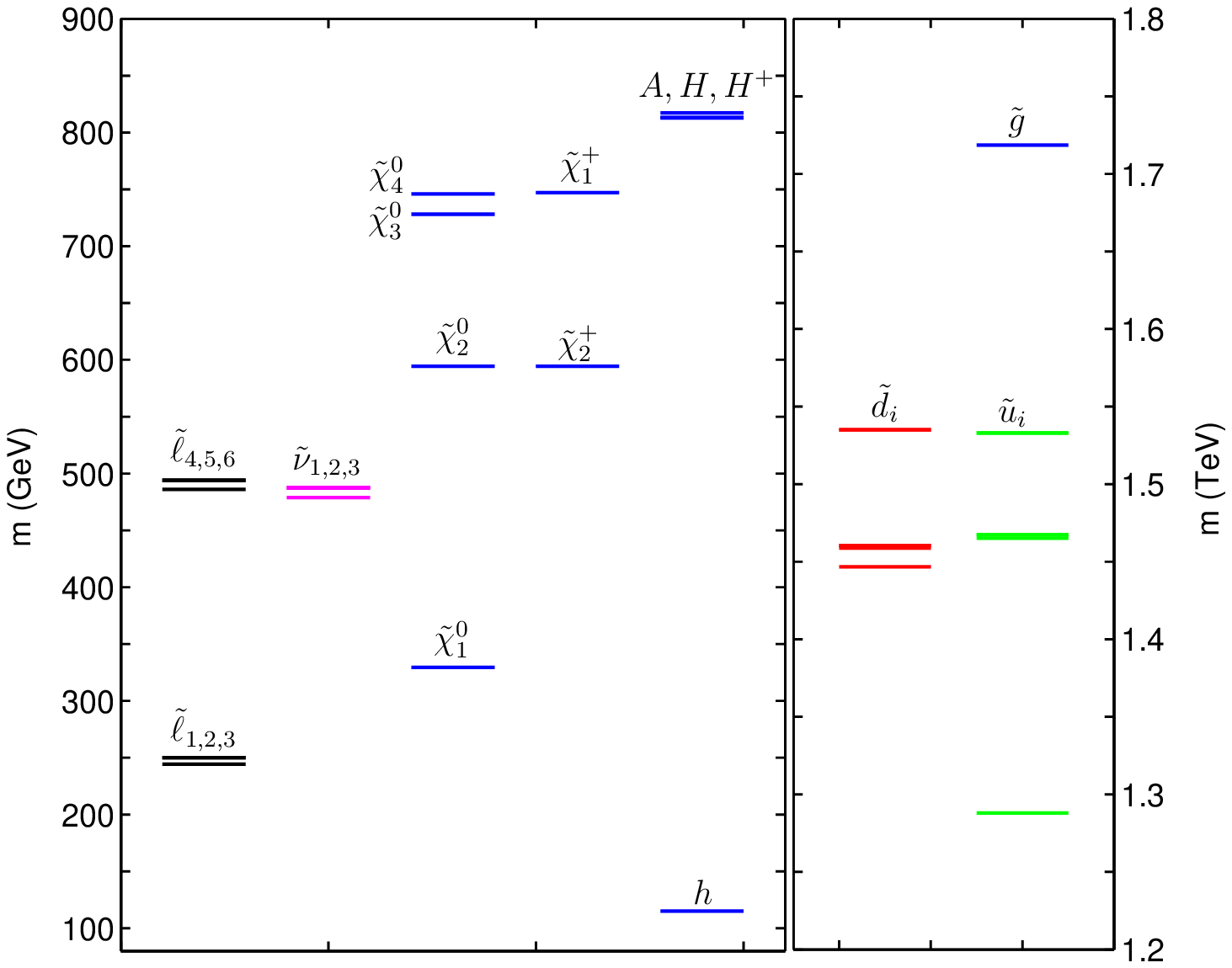}
\end{tabular}
\captions{\small Plots of the $N=4$ model for $B_T=60~{\rm TeV}$ and
normally ordered neutrino spectrum with $0=m_1<m_2 \ll m_3$ and
$s_{13}=0$. {\em Upper panels}: The $(\la_1, M_T)$ parameter space 
for $\mu M_a >0$ (left) and  $\mu M_a <0$ (right).
The pink region is excluded by the Higgs mass bound $m_h>$~114 GeV,
the grey one is excluded by perturbativity and the red one by the
EWSB conditions. Inside the yellow
area $\BR(\mu\to e \ga)$ is above the present experimental upper
bound. The dashed lines correspond to $\BR(\mu\to e \ga)=10^{-12},
10^{-13}, 10^{-14}$ (from top to bottom.). Isocontours of
$\tan\beta$ (solid/blue lines), $|\mu|$ (dash-dotted lines) and $\delta a_\mu$ (
solid/magenta lines) are also
shown. {\em Lower panel}: Sparticle and Higgs spectrum for $M_T=8
\times 10^7 \, \gev$ and $\la_1=0.25$. At this point of the
parameter space, which is marked by a star ($\star$) in the upper-right panel, 
$\tan\beta \simeq 11$, $\BR(\mu \rightarrow e \gamma) \simeq
1.6\times 10^{-12}$, $\BR(\tau \rightarrow \mu \gamma) \simeq 6 \times
10^{-10}$, $\BR(\tau \rightarrow e \gamma) \simeq 2 \times 10^{-13}$.} 
\label{f5}
\end{center}
\end{figure}

Some representative numerical results for the $N=4$ scenario are
shown in  Fig.~\ref{f5}. We set the scale of sparticle masses by
fixing $B_T=60~{\rm TeV}$, and consider a normally ordered neutrino
spectrum, with $0=m_1<m_2 \ll m_3$ and $s_{13}=0$. In the upper 
part of Fig.~\ref{f5} we show two plots of the $(\la_1,M_T)$ parameter
space, including contours of $\tan\beta$ and $\mu$ (extracted by
imposing EWSB). The left (right) panel corresponds 
to solutions of 
the EWSB conditions with $\mu M_a >0$ 
($\mu M_a <0$ ).
The main phenomenological
constraints come from the LFV decay $\mu \rightarrow e \gamma$ 
and 
the lightest Higgs mass\footnote{ We have included low-energy
corrections to $m_h$ by linking our code to {\em FeynHiggs}
\cite{feynh}.}. 
In both panels, the upper region is excluded by large values of
$\bY_T$, which either exceed the perturbative limits or generate
excessive LFV in $\bmm^2_{\tl}$, so that the bound $\BR(\mu
\rightarrow e \gamma) < 1.2 \times 10^{-11}$ is violated. The plots
also indicate other benchmark values of this BR, which will be
experimentally probed in the near future~\cite{MEG}. 
Other regions are excluded by the EWSB requirement.   
Notice that large values of $\tan\beta$ are achieved  for 
  $\mu M_a > 0$, which entails that in the corresponding parameter 
space the two tree-level contributions to the neutrino mass matrix
(\ref{mnusum}) have comparable size and opposite sign.  
An area in the
right part of the $\mu M_a < 0$ panel is excluded as well, since there the
tree-level contribution to $m_h$ is suppressed by low values of
$\tan\beta$, such that $m_h < 114 \, \gev$. Inside the allowed
portions of parameter space (shown in white), $m_h$ is around 
115 GeV.  We have also shown some contours of 
the SUSY contribution  $\delta a_\mu$
to the muon anomalous magnetic moment, 
which can have either sign in our model\footnote{
We recall that the discrepancy 
$\Delta a_\mu = a_\mu^{SM} - a_\mu^{\rm exp}$ between the 
SM prediction  and experiment has still some uncertainty, 
mainly related to the evaluation of the hadronic 
contribution to  $a_\mu^{SM}$. For instance, Refs.~\cite{gmuon} estimate 
$\Delta a_\mu \simeq - (2.5 \pm 0.8) \times 10^{-9}$ using $e^+ e^- $ data 
or $\Delta a_\mu \simeq - (1.6 \pm 0.8) \times 10^{-9}$ using $\tau$ data.}.
For $\mu M_a > 0$, the size of $\delta a_\mu$ and its positive sign  
are such that theory and experiment agree within $1 \sigma$.  
For  $\mu M_a < 0$ we have $\delta a_\mu < 0$, hence  
the discrepancy is not better than in the SM.
In this case we conservatively tolerate 
values of $\delta a_\mu$ up to $10^{-9}$ in magnitude.

In the lower panel of Fig.~\ref{f5} we show the sparticle and
Higgs spectrum for $M_T=8 \times 10^7 \, \gev$ and $\lambda_1=0.25$
(which corresponds to $\tan\beta \simeq 11$), again for $B_T=60~{\rm
TeV}$. The BRs of the LFV radiative decays are indicated
in the caption.  The Higgs sector is close to the decoupling limit,
since the states $A,H$ and $H^+$ are much heavier than $h$. Gluino
and squarks are the heaviest sparticles and the lightest of them is
${\tilde t}_1$ (which is mainly ${\tilde t}_R$). In the electroweak
sector, the heaviest chargino and neutralinos (${\tilde \chi}^+_2$,
${\tilde \chi}^0_{3,4}$) are mainly Higgsino-like, while ${\tilde
\chi}^+_1$ and ${\tilde \chi}^0_{2}$ are mostly Wino-like. The
(mainly left-handed) sleptons ${\tilde \ell}_{4,5,6}$ and the
sneutrinos ${\tilde \nu}_{1,2,3}$ are somewhat lighter than those
states, and the Bino-like neutralino ${\tilde \chi}^0_{1}$ is even
lighter. Finally, the lightest MSSM sparticles are the (mainly
right-handed) sleptons ${\tilde \ell}_{1,2,3}$, as generically
occurs in gauge mediated models with messenger index $N
>1$ and not too large mediation scale \cite{staunlsp}. The slepton
${\tilde \ell}_{1}$ (which is mainly $\tilde{\tau}_R$) is the
next-to-lightest SUSY particle (NLSP), while the gravitino ${\tilde
G}$ is the lightest SUSY particle (LSP). We recall that the latter
has mass $m_{\tilde G} = F/(\sqrt{3} M_P)$, where $M_P$ is the
Planck mass and $\sqrt{F}$ is the fundamental scale of SUSY
breaking.

The qualitative picture described above does not change very much under
variations of the model parameters. For instance, if we increase
$B_T$ the spectrum exhibits a roughly linear increase. The main
exception is $m_h$, which could increase by a few GeV, as a result
of the logarithmic corrections induced by larger stop masses.
Consequently, the rightmost boundary of the allowed parameter space
(upper-right panel of Fig.~\ref{f5}) would be shifted 
towards larger (smaller) values of $\lambda_1$ ($\tan\beta$). 
At the same time, the upper boundaries determined by 
the $\mu \to e \gamma$ constraint would slightly shift upwards, since a
heavier spectrum would imply smaller values for the LFV BRs.
Increasing  $B_T$ also reduces the magnitude of $\delta a_\mu$ 
($|\delta a_\mu| \propto 1/B_T^2$).  
Variations of $M_T$ induce logarithmic effects on the sparticle
spectrum. For low values of $M_T$, Higgsino and Wino
masses are closer to each other, and mixing effects in the chargino and
neutralino sectors are more important.
Moreover, the heavy Higgses can become lighter than one or both charginos.
This effect is more dramatic in the lower right corner of the
$\mu M_a >0$ parameter space, where those Higgs masses can decrease
even below 200 GeV.

The scenario described above can be tested at current and future
colliders. In particular, $p p$ collisions at the LHC should produce
a significant amount of squark pairs, either directly or through
associated squark/gluino production (followed by $\tilde{g}
\rightarrow \tilde{q} \bar{q}$) \cite{squarks}. For a spectrum as
the one shown in Fig.~\ref{f5} the production cross section is about
0.1 pb at $\sqrt{s}=14 \, \tev$. Once a $\tilde{q}$ is produced, it
can decay through well known chains, such as $\tilde{q}_R
\rightarrow q {\tilde \chi}^0_1 \rightarrow q \tau {\tilde \ell}_1$,
$\tilde{q}_L \rightarrow q {\tilde \chi}^0_2
 \rightarrow q \ell  {\tilde \ell} \rightarrow
q  \ell^+ \ell^- {\tilde \chi}^0_1  \rightarrow q  \ell^+ \ell^-
\tau {\tilde \ell}_1$, or similar ones with charginos and/or
sneutrinos (and neutrinos). Hence, in general, the final state of
such a $p p$ collision contains SM particles and two NLSPs ${\tilde
\ell}_{1}$, which eventually decay to $\tau {\tilde G}$ with rate
$\Gamma= m^5_{{\tilde \ell}_1}/(16 \pi F^2)$. The latter decay can
occur either promptly, or at a displaced vertex, or even outside the
main detector, as discussed in \cite{staunlsp,gmlhc}. Let us briefly
describe such possibilities in our case, taking into account that
$\sqrt{F} \gsim \sqrt{ \xi_T F_X } = \sqrt{B_T M_T}$ and $10^5 \,
\gev < \sqrt{B_T M_T} < 10^7 \, \gev$. {\it i}) For $\sqrt{F} < 10^7
\, \gev$, which includes the case $\sqrt{F}\sim \sqrt{ B_T M_T}$,
the NLSP decays occur inside the detector in most of our parameter
space, such that the escaping gravitinos contribute to the total
missing energy of the event. For instance, if a $p p$ collision
produces a pair of $\tilde{q}_R$, one can look for the overall
signature $p p \rightarrow  \tau^+ \tau^- \tau^+ \tau^- + 2 \, {\rm
jets} + {E}^{\rm miss}_T$, possibly with displaced vertices
corresponding to the NLSP decays. Instead, if a
$\tilde{q}_L$-$\tilde{q}_R$ pair is produced, the final state can
contain an additional lepton pair $\ell^+ \ell^-$. In both examples,
one more jet is present if one of the squarks originates from a
gluino. {\it ii}) For $\sqrt{F} \gsim 10^7 \, \gev$, each NLSP
${\tilde \ell}_1$ leaves a track in the main detector and mostly
decays outside. In the previous example with a $\tilde{q}_R$ pair,
the signature would be $p p \rightarrow 2 {\tilde \ell}_1 + 2 \tau +
2 \, {\rm jets}$. Moreover, in such cases the decay properties of
${\tilde \ell}_1$ could be measured by an additional massive
detector where ${\tilde \ell}_1$ may stop \cite{stausto}.



\subsection{Lepton flavour violation}

The experimental signatures mentioned above are also typical of a
class of gauge-mediated models, in which flavour is conserved in the
SUSY-breaking sector by construction. In contrast, in our scenario
LFV is intrinsically present and, therefore, LFV processes are a
crucial tool to discriminate our model from pure gauge mediation
ones. As already emphasized, the type II mechanism implies that LFV
is of minimal type and the basic flavour spurions are $\bY_T$ and
$\bY_e$. Let us focus, for simplicity, on the 
parameter space where  $\tan\beta$ is moderate. 
Then the leading LFV structure $\bY^\dagger_T
\bY_T$, which appears in $\bmm^2_{\tl}$ (and $\bA_e$), can be
related to the neutrino parameters as
\beq\label{rel1}
(\bmm^2_{\tl})_{ij}    \propto  B_T^2
(\bY^{\dagger}_T \bY_T)_{ij}  \propto
\left(\frac{M_T^2 \tan^2 \! \beta }{\la_1}\right)^2
 \left[\bV (\bmm^{D }_\nu)^2 \bV^\dagger\right]_{ij} \propto
\tan^5 \! \beta \, M_T^4
 \left[\bV (\bmm^{D }_\nu)^2 \bV^\dagger\right]_{ij}\,,
\eeq
where in the last step we have traded the $\la_1$-dependence for a
$\tan\beta$-dependence through the EWSB 
conditions, {\it i.e.} $\la_1^2 \sim B_H \sim 1/\tan\beta$
(approximatively valid for $\tan\beta \lsim 20$). Notice that
eq.~(\ref{rel1}) includes several approximations, while in the
numerical analysis the matching between $\bY_T$ and $\bmm_\nu$
proceeds through the effective operators and takes into account RGE
effects. However, the latter do not introduce any unknown flavour
structure [see eq.~(\ref{rgek})] and thus the structure of $\bY_T$
at the scale $M_T$ can be unambiguously predicted (modulo an overall
unflavoured factor) once the observable $\bmm_\nu$ is experimentally
determined~\cite{AR,JR,FJ}.

As a consequence of the misalignment between slepton and lepton mass
matrices, LFV signals can appear either at high-energy colliders or
in low-energy processes. Concerning the former possibility, LFV
could show up at the LHC in, \eg , neutralino decays, such as
${\tilde \chi}^0_2 \rightarrow \ell_i^{\pm} {\tilde \ell}_a^{\mp}
\rightarrow \ell_i^{\pm} \ell_j^{\mp} {\tilde \chi}^0_1$ with $i
\not= j$ \cite{neulfv,lfvcern,porod}, or ${\tilde \chi}^0_1
\rightarrow \ell^{\pm} {\tilde \ell}_1^{\mp}$ with $\ell \not= \tau$
\cite{k08}. Moreover, since LFV also affects
lepton-slepton-gravitino couplings \cite{br00}, a small fraction of
NLSP decays ${\tilde \ell}_1 \rightarrow \ell {\tilde G}$ could
produce a lepton $\ell \not= \tau$ \cite{staulfv}. As LFV appears
mostly in the left sector in our scenario, the relevant LFV channel
in the above examples is expected to be ${\tilde \chi}^0_2
\rightarrow \ell_i^{\pm} \ell_j^{\mp} {\tilde \chi}^0_1$, while LFV
effects in the subsequent decays of ${\tilde \chi}^0_1$ and ${\tilde
\ell}_1$ are more suppressed since ${\tilde \ell}_1\sim {\tilde
\tau}_R$. For instance, one could look for the LFV decays ${\tilde
\chi}^0_2 \rightarrow \mu^{\pm} \tau^{\mp} {\tilde \chi}^0_1$ or
${\tilde \chi}^0_2 \rightarrow \mu^{\pm} e^{\mp} {\tilde \chi}^0_1$,
followed by the flavour-conserving decay ${\tilde
\chi}^0_1\rightarrow \tau^+ \tau^- {\tilde G}$ (or ${\tilde
\chi}^0_1  \rightarrow {\tilde \ell}_1^{\pm} \tau^{\mp}$ if the NLSP
is long-lived). These options, and perhaps others as well, may
deserve further studies.

We now focus on the radiative decays $\ell_i \to \ell_j \ga $ (for
which we have already shown some predictions in Fig.~\ref{f5}),
taking them as representative low-energy LFV processes. By using
eq.~(\ref{rel1}), we can infer that
\beq
\label{brdep}
\BR(\ell_i \to \ell_j \ga) \propto \left[ \frac{(\bmm^2_{\tl})_{ij}} { \tm^4 }
 \tan\!\beta \right]^2
\propto \left(\frac{M_T}{B_T}\right)^{\!8 }\!(\tan\beta)^{12}
\left[\bV (\bmm^{D }_\nu)^2 \bV^\dagger\right]_{ij}\,.
\eeq
This approximate relation expresses the $\BR$s as a product of a
common overall factor (which depends on high powers of $M_T$, $B_T$
and $\tan\beta$) and a flavour-dependent one, which is determined by
the low-energy neutrino parameters only. The latter property is
typical of type II models \cite{AR, JR}, while the form of the
overall `unflavoured' factor depends on the specific realization. If
we take ratios of $\BR$s, the overall factor drops out and we obtain
the following estimates\footnote{ We have inserted the best fit
values of the neutrino parameters quoted above and assumed a
normally ordered neutrino spectrum. The results for inverted
ordering are obtained by exchanging $\delta =0 \leftrightarrow
\delta =\pi$. Similar results hold when the quartic terms
$(\bY^\dagger_T \bY_T)^2$ dominate over the quadratic ones
$\bY^\dagger_T \bY_T$ in $\bmm^2_{\tl}$ [see eq.~(\ref{mlsoft})].
The results change if cancellations occur between such terms. A
detailed analysis of this case could be performed as in \cite{JR},
where similar effects are present.}:
\bea
\label{brs1}%
\frac{\BR(\tau \to \mu\ga)}{\BR(\mu \to e\ga)} & \approx &
\left[\frac{(\bmm^2_{\tl})_{ \tau \mu}}{(\bmm^2_{\tl})_{\mu
e}}\right]^2 \frac{\BR(\tau \to \mu \nu_\tau \bar{\nu}_\mu)}
{\BR(\mu \to e \nu_\mu \bar{\nu}_e)} \approx\left\{\begin{array}{l}
400\quad\quad\quad \quad [s_{13}=0 \, ] \\
\\
2 ~(3)\!\!\quad\quad\quad \quad  [s_{13}=0.2\, , \;  \delta = 0 \;
(\pi)]
\end{array} \right. \no  \\
&& \no \\
&& \no \\
\frac{\BR(\tau \to e\ga)}{\BR(\mu \to e\ga)} & \approx &
\left[\frac{(\bmm^2_{\tl})_{ \tau e}}{(\bmm^2_{\tl})_{\mu
e}}\right]^2 \frac{\BR(\tau \to e \nu_\tau \bar{\nu}_e)} {\BR(\mu
\to e \nu_\mu \bar{\nu}_e)} \approx \left\{\begin{array}{l}
0.2\quad\quad\quad \quad \,\,\,\, [s_{13}=0 \,] \\
\\
0.1 ~(0.3)\,\,\quad \quad  [s_{13}=0.2\, , \; \delta = 0 \;
(\pi)]\,.
\end{array}\right.
\no \\
&&
\eea
We recall that the approximate relations (\ref{rel1}), 
(\ref{brdep}) and (\ref{brs1}) hold for small or moderate $\tan\beta$.
We will not discuss the special features  that emerge for large
$\tan\beta$, which could be analysed, \eg, along the lines of~\cite{FJ}.

\begin{figure}[h]
\begin{center}
\begin{tabular}{ll}
\hspace*{-0.4cm}
\includegraphics[width=8cm]{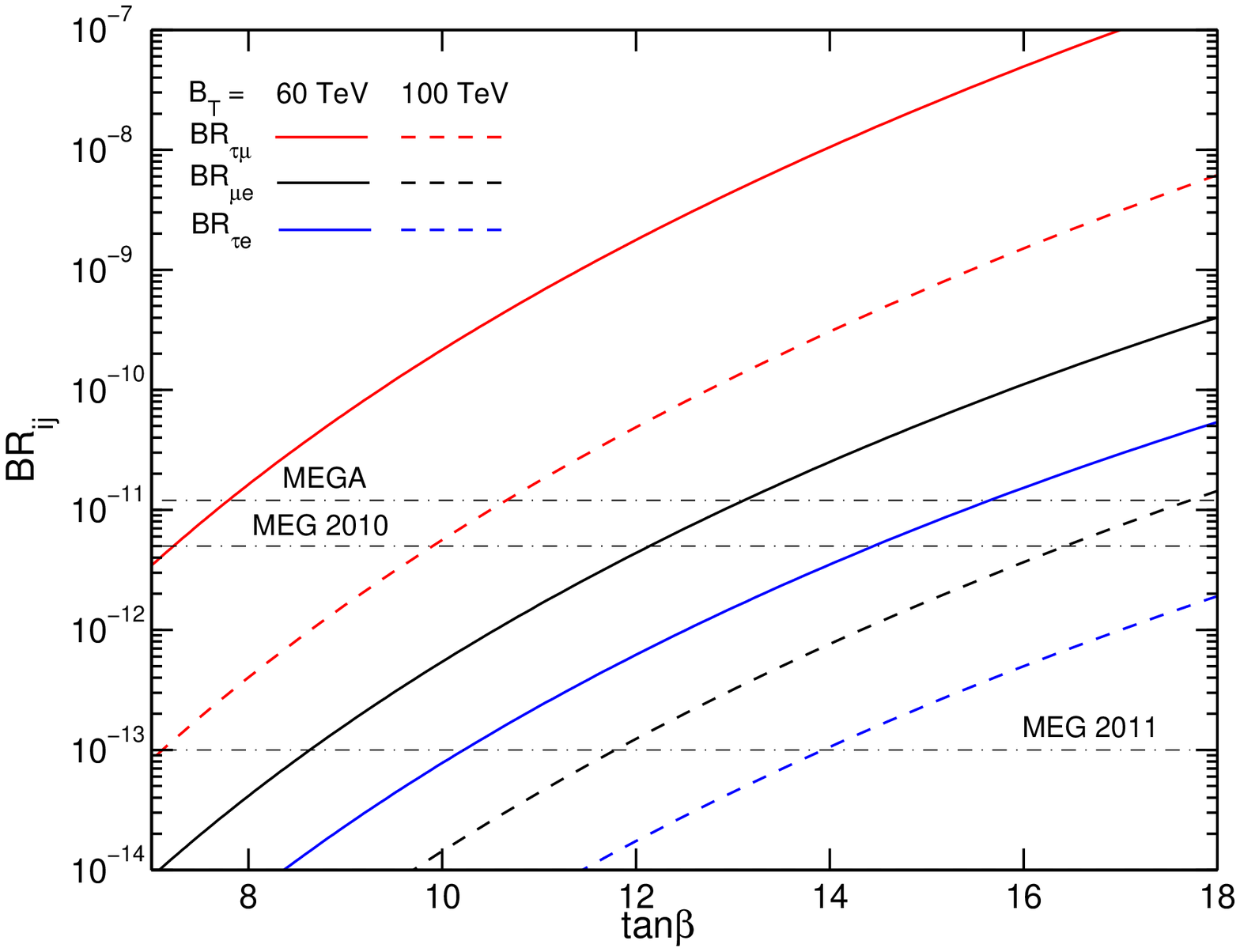} &
\hspace*{-0.4cm}\includegraphics[width=8.3cm]{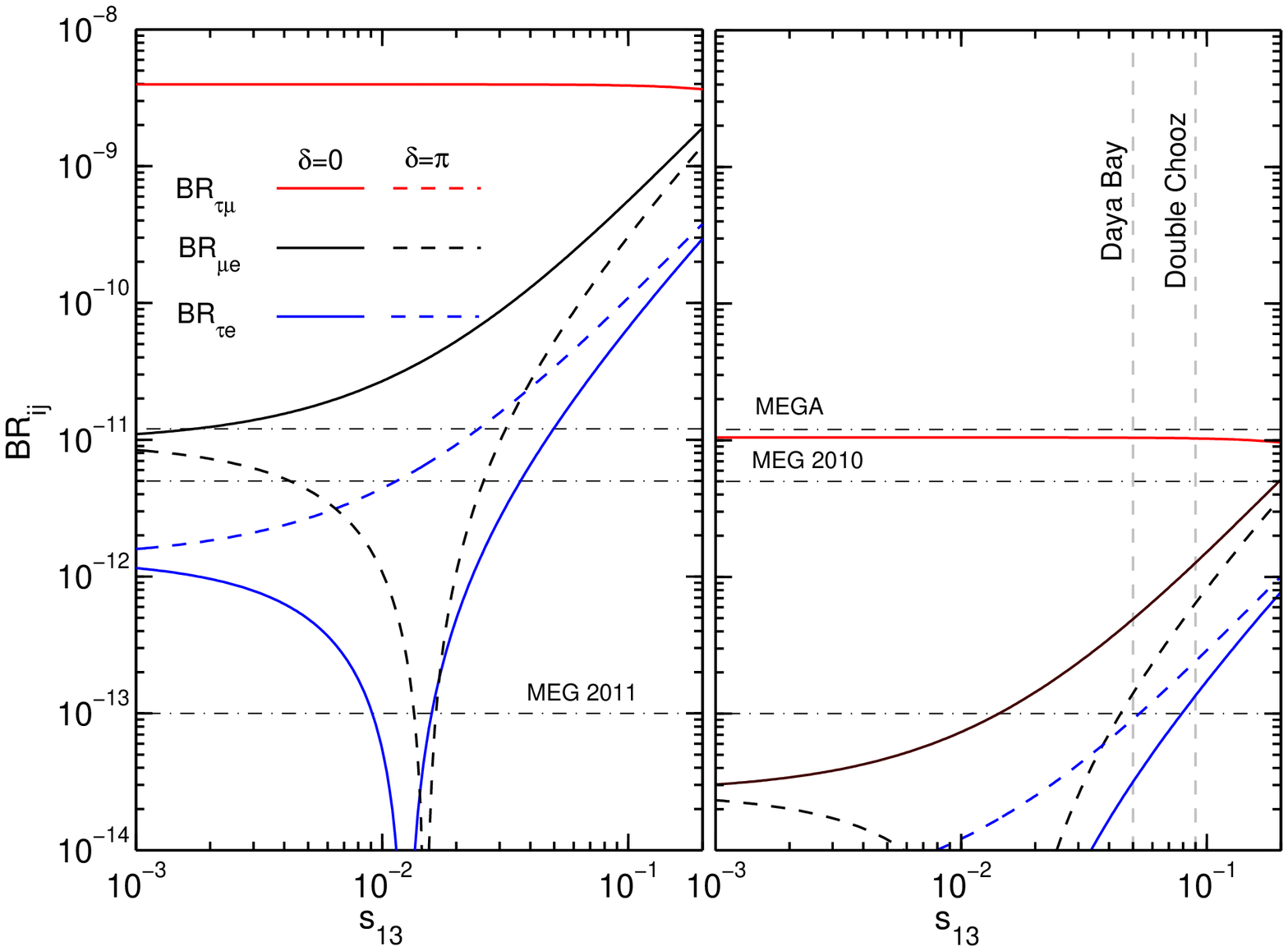}
\end{tabular}
\captions{\small Plots of $\BR(\ell_i \to \ell_j \gamma)$ for $M_T =
8\times 10^7\,{\rm GeV}$. The present (MEGA) and near-future (MEG)
sensitivities  on $\BR(\mu \to e \gamma)$ are also shown. {\it Left
panel}: The BRs as a function of $\tan\beta$ for $B_T= 60\,{\rm
TeV}$ (solid lines) or $B_T= 100\,{\rm TeV}$ (dashed lines), with
$s_{13}=0$. {\it Right panel}: The BRs as a function of $s_{13}$ for
$B_T= 60\,{\rm TeV}$, with $\tan\beta = 13 \, (8)$ in the first
(second) subpanel. For $\BR(\mu \to e \gamma)$ and $\BR(\tau \to e
\gamma)$, the solid (dashed) curves correspond to $\delta = 0 \,
(\pi)$ assuming a normally-ordered neutrino spectrum. We also show
the future sensitivity for $s_{13}$ (the RENO expected sensitivity
lies between those of Daya Bay and Double Chooz). } \label{f6}
\end{center}
\end{figure}

Let us now present some numerical examples. In the left panel of
Fig.~\ref{f6} the three $\BR(\ell_i \to \ell_j \ga)$ are shown as a
function of $\tan\beta$, taking either $B_T= 60\,{\rm TeV}$ (solid
lines) or $B_T= 100\,{\rm TeV}$ (dashed lines), for $M_T = 8\times
10^7\,{\rm GeV}$ (implying $\mu M_a <0$), normally ordered neutrino spectrum and $s_{13}=0$.
Notice that we have traded $\la_1$ for $\tan\beta$, which is a more
physical and testable parameter. The behaviour of the $\BR$s 
and their mutual ratios are 
consistent with the qualitative predictions of eqs.~(\ref{brdep}) and 
(\ref{brs1}).  For
$B_T= 60\,{\rm TeV}$, which corresponds to a sparticle spectrum
testable at the LHC (see Fig.~\ref{f5}), $\BR(\mu \to e \ga)$ can be
tested at the MEG experiment~\cite{MEG} if $\tan\beta \gsim 7$. For
$B_T = 100\,{\rm TeV}$, gluino and squark masses are pushed above
2~TeV, so the discovery of SUSY at the LHC would require more
integrated luminosity than in the previous example. In this case LFV
decays can still be a valid probe of our scenario, since $\BR(\mu
\to e \ga)$ could be discovered by MEG for $\tan\beta \gsim 10$.
Obviously, the ranges of $\tan\beta$ quoted in these examples would
change under variations of $M_T$ and $B_T$, as can be inferred from
eq.~(\ref{brdep}) (see also Fig.~\ref{f5}). Also, from both cases we
can see that if $\BR(\mu \to e \ga)$ is close to its present bound,
$\BR(\tau \to \mu \ga)$ is above $10^{-9}$, within the reach of
future Super Flavour Factories \cite{KEKB,SFF}.

The double panel on the right of Fig.~\ref{f6} illustrates the
dependence of the BRs on the least known neutrino parameters, namely
$s_{13}$ and $\delta$, for $B_T= 60\,{\rm TeV}$ and $M_T = 8\times
10^7\,{\rm GeV}$, with $\tan\beta = 13 \, (8)$ in the first
(second) subpanel. Regarding $\BR(\mu \to e \gamma)$ and $\BR(\tau
\to e \gamma)$, the solid (dashed) curves correspond to $\delta = 0
\, (\pi)$, while the region between such curves is spanned by
intermediate values of $\delta$. The dependence of $\BR(\tau \to \mu
\gamma)$ on $s_{13}$ and $\delta$ is negligible. The first subpanel
corresponds to a scenario which could be tested very soon at MEG
through the search of $\mu \to e \gamma$, if $s_{13} \ll 0.01$.
Notice that $\BR(\tau \to \mu \gamma)$ in this example is around $4
\times 10^{-9}$, within the reach of future Super Flavour Factories
\cite{KEKB,SFF}, while $\tau \to e \gamma$ would be unobservable
because $\BR(\tau \to e \gamma)\sim 10^{-12}$. For $s_{13} \sim
0.01$, $\BR(\mu \to e \gamma)$ and $\BR(\tau \to e \gamma)$ can be
either enhanced or suppressed since, depending on the value of
$\delta$, a cancellation can occur in the LFV quantity $\left[\bV
(\bmm^{D }_\nu)^2 \bV^\dagger\right]_{ij}$ \cite{JR,FJ} (see also
the third ref. in \cite{mfv}). The cancellation takes place in
$\BR(\mu \to e \gamma)$ [$\BR(\tau \to e \gamma)$] for $\delta = \pi
\, (0)$ in the case of normal ordering, while the opposite occurs
for inverted ordering. If $\BR(\mu \to e \gamma)$ is suppressed by
that cancellation mechanism, only $\tau \to \mu \gamma$ can be
observed. In such a case,  we can even obtain values of $\BR(\tau
\to \mu \gamma)$ above $ 10^{-8}$ (\ie, close to its present bound
\cite{babar,belle})
 by slightly changing the model parameters [see
eq.~(\ref{brdep})]. In the case of partial cancellations, $\mu \to e
\gamma$ could be still probed by MEG for values of $s_{13}$ up to
about 0.03, which are in the potential reach of future Neutrino
Factories \cite{NF}. The second subpanel shows an alternative
possibility, in which LFV $\tau$ decays are invisible, whereas
$\BR(\mu \to e \gamma)$ lies in the range $10^{-13} - 5 \times
10^{-12}$ if $0.05 \lsim s_{13} < 0.2$. Those values of $\BR(\mu \to
e \gamma)$ should be probed by MEG next year, while the indicated
range of $s_{13}$ is within the sensitivity of the present
accelerator experiments MINOS \cite{minos}, OPERA \cite{opera} and
T2K \cite{T2K}, and the incoming one NOvA \cite{NOvA}, as well as of
the near-future reactor experiments Double Chooz \cite{DC}, Daya Bay
\cite{DB} and RENO \cite{RENO}. This example shows the importance of
the interplay between LFV searches and neutrino oscillation
experiments.

For the sake of completeness, we recall that there are good
prospects to observe $\mu$-$e$ LFV also through the processes $\mu
\to e e e$ and $\mu \to e$ conversion in nuclei, whose rates are
correlated to $\BR(\mu \to e \gamma)$  in our scenario. In
particular, dipole-operator dominance implies $\CR(\mu \to e; {\rm
Ti}) \simeq 5 \times 10^{-3}\, \BR(\mu \to e \gamma)$ and 
 $\CR(\mu \to e; {\rm
Al}) \simeq 3 \times 10^{-3}\, \BR(\mu \to e \gamma)$
\cite{hisano,JR}. Hence, if MEG discovers $\mu \to e \gamma$ then
$\mu \to e$ conversion could be tested by the dedicated experiments
planned at J-PARC \cite{comet} and Fermilab \cite{mu2e}.


\section{Conclusions}

In the last decade the flourish of experimental
neutrino data has provided robust evidence of non-vanishing neutrino masses and
 mixing angles. This has stimulated further efforts and ideas
to understand the origin and pattern of neutrino masses. This issue
is part of the wider SM `flavour problem'. In the case of neutrino
masses, one should explain both their flavour structure and their
suppression with respect to the charged fermion masses. In this work
we have especially addressed the latter aspect in a SUSY framework.
At variance with the standard approach, in which neutrino masses
effectively arise from the $d=5$, $\Delta L=2$ superpotential
operator $(H_2 L)^2/M$, we have focussed on an interesting
alternative mechanism, which relies instead on the $d = 6$, $\Delta
L=2$ K\"{a}hler operator $(H^\dagger_1 L)^2/M^2$, previously
proposed in \cite{CEN}. We have discussed and further elaborated
this idea, first giving a comprehensive model-independent
description  in an effective-theory approach and then presenting
explicit realizations.

In particular, in Section~2 we have investigated the above effective
operator together with three novel related ones which emerge from
SUSY-breaking insertions. In principle, these four $\Delta L=2$
operators have independent coefficients and flavour structures. Two
of them contribute to neutrino masses at the tree level and the
other two through one-loop corrections at the sparticle threshold.
We have also computed the full set of one-loop RGEs, which are
required to relate the low-energy effects of those operators with
the high-energy scale where they emerge. The effective-theory
description we have presented holds both in the MSSM and in simple
extensions such as the NMSSM.

In Sections~3 and 4 we have proposed  a simple explicit realization
of those K\"{a}hler operators in a type II seesaw framework, namely
by the exchange of heavy $SU(2)_W$-triplet states. The SUSY operator
emerges at the tree level, while the origin and the size of the
SUSY-breaking ones depend on the mechanism and the scale $M_S$ of
SUSY-breaking mediation. In particular, the coefficients of the
latter operators are related to the SUSY-breaking parameters  of the
triplet states if $M_S \geq M_T$, while they can be induced
radiatively if $M_S < M_T$. Finally, we have focussed on the special
case $M_S = M_T$ and proposed a predictive scenario in which the
triplets are messengers of both lepton-number violation and SUSY
breaking, as in \cite{JR}.

In the case with $M_S=M_T$, the MSSM sparticle masses arise by
triplet-exchange at the quantum level, via both gauge and Yukawa
interactions. In order to generate a mass for the gluino and to
preserve perturbative gauge coupling unification, we have embedded
the triplets in a messenger sector with coloured states. The free
parameters of our model are only three, namely the triplet mass
$M_T$, the effective SUSY-breaking scale $B_T$ and the dimensionless
coupling $\lambda_1$ (which can be traded for $\tan\beta$ through
the EWSB condition). The messenger index
$N$ is an additional discrete parameter lying between 4 and 7.
Correlations exist among several observables, such as neutrino
parameters, sparticle and Higgs masses, and LFV decay rates. A
numerical analysis of the parameter space, for the minimal value
$N=4$, reveals that the model is phenomenologically viable for $B_T
> 50 \, \tev$ and $10^5\,\gev < M_T < 10^9\, \gev$. The latter range
is reduced for larger values of $N$, since the lower bound on $M_T$
increases.

The MSSM sparticle spectrum is analogous to that of pure gauge
mediation models with $N>1$ and not too large mediation scale. The
heaviest MSSM sparticle is the gluino, while the lightest one is a
stau. The latter, which is in fact the NLSP, can be either short or
long-lived and decays into $\tau$ and gravitino. As far as $B_T
\lsim 100\,\tev$, the sparticle spectrum can be probed at the LHC
(see Fig.~5) through the production of squarks and gluinos and their
subsequent decays into the remaining sparticles.

The presence of LFV allows us to distinguish our scenario from pure
gauge mediation models. Such a feature can be tested through the
search of either LFV sparticle decays or low-energy LFV processes.
We have also emphasized that LFV is of minimal type, as always
occurs in type II realizations of the seesaw mechanism. In
particular, the flavour structure of the slepton mass matrix is
essentially determined by the low-energy neutrino parameters. As a
result, the ratios of the $\BR(\ell_i \to \ell_j \gamma)$ can be
determined in terms of those parameters,
while the absolute values of the BRs depend also on $M_T$, $B_T$ and
$\tan\beta$.
In particular,  we can envisage several scenarios for  the detection
of LFV signals, depending on the yet unknown parameter $s_ {13}$.
 {\it i})~If $s_ {13}\ll 0.01$ (beyond the planned experimental sensitivity),
there are portions of the parameter space in which $\BR(\mu \to e
\gamma)$ and   $\BR(\tau \to \mu \gamma)$ are in the reach of the
MEG experiment and Super Flavour Factories, respectively. {\it
ii})~If $s_ {13}\sim 0.01$ (in the potential reach of future
Neutrino Factories) and for suitable values of $\delta$ [\ie\,
$\delta \sim \pi\, (0)$ for normal (inverted) ordering in the
neutrino spectrum],
 $\BR(\mu \to e \gamma)$ is strongly suppressed while  $\BR(\tau \to \mu \gamma)$
can be experimentally accessible. {\it iii})~ If $ s_{13} \gsim 0.1$
(reachable by near-future reactor and accelerator neutrino
experiments), only $\mu$-$e$ LFV can be probed through the
measurement of $\BR(\mu \to e \gamma)$ by the MEG collaboration.

In conclusion, we have discussed a SUSY scenario which provides an
alternative explanation for the smallness of neutrino masses and
relates them to sparticle masses in a specific type II realization.
Along these lines, further investigations can be envisaged. For
instance, on the theoretical side one could address the other aspect
of the `flavour problem', which concerns the origin of the flavour
structure of $\bmm_\nu$ (or $\bY_T$), and explore possible
connections with grand unified theories. On the phenomenological
side, one could study in more detail the correlation between the
sparticle spectrum and the LFV signals, by taking advantage of the
interplay between the LHC and low-energy experiments. This would
possibly help to discriminate among different mechanisms of
SUSY-breaking mediation in the context of the type II seesaw.

\vglue 2cm


\end{document}